\documentclass[fleqn,usenatbib]{mnras}

\usepackage{newtxtext,newtxmath}

\usepackage[T1]{fontenc}

\DeclareRobustCommand{\VAN}[3]{#2}
\let\VANthebibliography\thebibliography
\def\thebibliography{\DeclareRobustCommand{\VAN}[3]{##3}\VANthebibliography}

\usepackage{graphicx}	
\usepackage{amsmath}	

\usepackage{amssymb}	
\usepackage{bbold}      
\usepackage[normalem]{ulem}       
\usepackage{commath}
\usepackage{bm}

\title[First 3-D grid-based gas-dust simulations]{First 3-D grid-based gas-dust simulations of circumstellar disks with an embedded planet}

\author[F. Binkert et al.]{
Fabian Binkert,$^{1,2,3}$\thanks{E-mail: fbinkert@usm.lmu.de}
Judit Szul\'{a}gyi$^{3,4}$
and Til Birnstiel$^{1,2}$
\\
$^{1}$University Observatory, Faculty of Physics, Ludwig-Maximilians-Universität München, Scheinerstr. 1, 81679 Munich, Germany\\
$^{2}$Exzellenzcluster ORIGINS, Boltzmannstr. 2, D-85748 Garching, Germany\\
$^{3}$Institute for Particle Physics \& Astrophysics, ETH Zurich, Wolfgang-Pauli-Str. 27, 8093 Zürich, Switzerland\\
$^{4}$Institute for Computational Science, University of Zurich, Winterthurerstrasse 190, 8057 Zürich, Switzerland
}

\date{Accepted XXX. Received YYY; in original form ZZZ}

\pubyear{2020}

\begin{document}
\label{firstpage}
\pagerange{\pageref{firstpage}--\pageref{lastpage}}
\maketitle

\begin{abstract}
Substructures are ubiquitous in high resolution (sub-)millimeter continuum observations of circumstellar disks. They are possibly caused by forming planets embedded in {their} disk. To investigate the relation between observed substructures and young planets, we perform novel three-dimensional two-fluid (gas+1-mm-dust) hydrodynamic simulations of circumstellar disks with embedded planets (Neptune-, Saturn-, Jupiter-, 5 Jupiter-mass) at different orbital distances from the star (5.2AU, 30AU, 50AU). We turn these simulations into synthetic (sub-)millimeter ALMA images. We find that all but the Neptune-mass planet open annular gaps in both the gas and the dust component of the disk. We find that the temporal evolution of the dust density distribution is distinctly different {from} the gas’. For example, the planets cause significant vertical stirring of the dust in the circumstellar disk which opposes the vertical settling. This creates a thicker dust disk than disks without a planet. We find that this effect greatly influences the dust masses derived from the synthetic ALMA images. Comparing the dust disk masses in the 3D simulations and the ones derived from the 2D ALMA synthetic images, we find the former to be a factor of a few (up to 10) larger, pointing to that real disks might be significantly more massive than previously thought based on ALMA continuum images using the optically thin assumption and equation. Finally, we analyze the synthetic ALMA images and provide an empirical relationship between the planet mass and the width of the gap in the ALMA images, including the effects of the beam size.

\end{abstract}




\section{Introduction}
Substantial theoretical work on the interaction between young planets and their hosting circumstellar disk has been carried out over the past decades. Both analytical and numerical approaches have lead to an improved understanding of the problem \citep[e.g.][]{GoldreichTremaine1980,LinPapa1986,Tanaka2002,Paardekooper2004,Paardekooper2006,DeVal-Borro2006,Kley2012}. 
Today, it is widely accepted that a young planet embedded in a circumstellar disk can open one or even multiple annular gaps and/or rings in the gaseous and dusty components of the disk \citep[e.g.][]{Dong2015,Picogna2015,Jin2016,Fedele2017,Bae2017}. However, it is still up for debate whether the annular structures, seen in scattered light and mm-continuum observations of circumstellar disks, are indeed of planetary origin or if they have formed via other processes, e.g. dust pile-up at condensation fronts \citep{Zhang2015}, dead zones \citep{Ruge2016} or large-scale vortices \citep{Barge2017}. However, the planetary hypothesis is obvious because planets are found to be common around stars. Yet, it remains difficult to directly observe young planets embedded in the hosting circumstellar disk. Only a few planetary candidates still embedded in their circumstellar disk have been observed, e.g. PDS 70 b, c \citep{Mueller2018,Keppler2018}. Therefore, studies have focused on observable disk substructures to indirectly probe the properties of the unseen planet population and to establish a link between observations and planet formation theory. Current efforts mainly focus on near-IR scattered light images \citep{Dong2015,Avenhaus2018,Szulagyi2019}, (sub-)mm dust thermal continuum images \citep{Zhang2018,Szulagyi2018b}, or molecular line channel maps which trace the gas kinematics \citep{Perez2015,Pinte2018,Teague2018}. \newline
The planetary gaps in disks exist due to the exchange of angular momentum between a planet and the surrounding disk. The radial distribution of the disk material adjusts due to the gravitational transfer of angular momentum {i.e. via gravitational torques} from the inner part of the disk to the planet and from the planet to the outer part of the disk\citep{GoldreichTremaine1980,Lin1984}. Hence, disk material is radially pushed away from the planet. In gas, a gap opens if the gravitational torques win over the counterbalancing viscous torques. In addition to these two torques, the gas near the planet also feels a torque due to pressure due to the non-axisymmetric nature of the planetary wakes \citep[][]{Crida2005}. In a steady state, viscous torques, gravitational torques caused by the planet and torques due to pressure all balance each other. {Hence, gap opening in gas does, in addition to its dependence on the planetary mass, depend on disk properties such as temperature \citep{Crida2005,Szulagyi2017,ZhangZhu2020,Ziampras2020} and turbulent viscosity \citep{Lin1993}.} In other words, gap opening is more efficient, i.e. leads to deeper and wider gaps, for more massive planets in low viscosity gas with a larger Mach number \citep{Crida2005}. \newline
In addition to gas, circumstellar disks consist to about 1\% of their mass of solid material. In early phases, most of the solid material is present in the form of dust, i.e. solid grains of sizes ranging from below micron up to a few centimeters. These particles are suspended in the mass dominating gaseous component of the disk and are coupled to the gas via aerodynamic drag forces. Drag forces and gravitational forces of the central star cause the dust grains to radially drift and vertically settle. In addition to these two transport mechanisms, dust grains experience turbulent diffusion due to their coupling to the gas' turbulent motion. Turbulent diffusion smooths out gradients in the local dust-to-gas ratio and, thus, can counteract radial drift or vertical settling. The result is a finite vertical thickness of the dust layer and the absence of sharp features in the dust distribution. Generally, turbulent diffusion is the dominant process for small dust grains whereas radial drift and vertical settling is more dominant for large dust grains. \citep[][]{Dubrulle1995,Youdin2007} \newline
{Due to its different nature, gap}  opening in dust is somewhat different {from gap opening in gas}. \cite{Dipierro2016} differentiate two mechanisms. Small dust particles, which strongly couple to the gas, move along with the gas and therefore show similar gap opening characteristics as the gas. Larger dust grains, however, decouple from the gas and drift in the direction of the disk pressure gradient. If a planet has already opened a gap in the gas, weakly coupled dust grains accumulate at the pressure maxima located at the gap edges. The result is a depletion of larger dust grains in the gap region due to radial drift \citep{Paardekooper2004,Paardekooper2006,Fouchet2007,Fouchet2010}. However, the presence of a gap in the gas is not a necessary condition for gap opening in dust. For low-mass planets, a gap can be present in the dust only \citep{Dipierro2017}. Similar to the case in gas, gap opening in dust also occurs when gravitational torques push away dust from the planetary orbit. However, unlike in gas, viscous torques are not present in dust when it is treated as a pressureless inviscid fluid. In a disk without a planet, the aerodynamic drag torque acting on dust particles due to the interaction with the gas is in general negative, resulting in a radial inward drift \citep{Nakagawa1986}. When a planet is present, gravitational torques also contribute. In the inner disk (inside the planetary orbital radius), gravity and aerodynamic torques add up and lead to an inward drift of the dust. In the outer disk, the two torques counteract each other. Hence, if the gravity torque is strong enough, i.e. the planet is massive enough, it can prevent dust from drifting from the outer disk to the inner disk and a gap opens in the dust without the need for a pressure bump in the gas \citep{Johansen2009}. Therefore, gap opening in dust depends, besides on the mass of the planet, on the degree of coupling between dust and gas, and also on the size of the dust particles. This has been shown in various numerical studies  \citep[e.g.][]{Paardekooper2004,Paardekooper2006,Fouchet2007,Fouchet2010,Rosotti2016}. \newline
Planetary gaps and rings are predicted to be detectable in (sub-)mm continuum observations using the latest generation of radio interferometers such as the Atacama Large Millimeter/submillimeter Array (ALMA) \citep[e.g.][]{Pinilla2012,Gonzalez2012,Gonzalez2015,Pineda2019}. More recently, high-resolution ALMA observations have indeed revealed a multitude of substructures such as gaps, rings, spirals, and large-scale asymmetries in circumstellar disks, e.g. \cite{vandermarel2013,ALMApartnership2015,Andrews2018}. Even though the planetary origin of the observed features is still being debated, and young planets remain difficult to detect, the observed substructures can be used to indirectly probe the properties of the unseen population of forming planets. Continuum observations at (sub-)mm-wavelength most efficiently probe the thermal emission of mm-sized dust grains \citep{Draine2006} coming from the cold midplane region of a circumstellar disk. In this region, even low-mass {giant} planets can cause substructures in mm-sized dust. {Typically, the lower mass limit for a planet to open a gap in 1 mm-sized dust is on the order of a Neptune-mass \citep[e.g][]{Paardekooper2006,Fouchet2007}} \newline
A lot of work has been put into making observational planet-disk-interaction predictions for ALMA based on hydrodynamic models \citep[e.g.][]{Gonzalez2012,Ruge2016,Szulagyi2018a,Zhang2018,IsellaTurner2018,Dipierro2018}. In these models, an accurate dynamic and thermal treatment of the dust component in the disk is crucial to produce realistic observational predictions in (sub-)mm continuum observations. Moreover, radiative transfer methods are necessary to create synthetic ALMA images because there is no one-to-one relation between hydrodynamic features and observed features. Thermal emissions are dependent on a combination of density, temperature, and optical properties of the emitting region. An accurate computational treatment of all these quantities is therefore crucial for making accurate observational predictions. However, most studies compromise on physical accuracy in favor of computational efficiency. \newline
In this work, we aim to improve upon some of the shortcomings of previous studies and create physically accurate observational predictions for planet-induced substructures in circumstellar disks. Rather than using a particle-based approach for the dust component in the hydrodynamical models, we use a grid-based approach for both the gas and dust components. A grid-based approach does not suffer from a lack of resolution in low-density regions. Moreover, we perform global hydrodynamical simulations in three dimensions instead of two dimensions because planet-disk interaction is inherently a three-dimensional problem. We also avoid the common isothermal assumption in gas and include heating (adiabatic heating, viscous heating, stellar irradiation) and cooling processes (adiabatic{, radiative} cooling) in our thermal simulations. To our knowledge, no global three-dimensional grid-based dust and gas planet-disk-interaction simulations have yet been published. We carry out a total of 12 hydrodynamic simulations using two fluids (gas + mm-sized dust) in which we embed a planet in the disk. We use different planetary parameters (mass, orbital radius) in each simulation and produce synthetic mm-continuum observations for ALMA with realistic beam-sizes from the hydrodynamical models. This allows us to study the observable disk features in the mm-continuum induced by the planets. We derive an empirical formula that relates the planet mass to the width of the observed gaps. We also compare total dust masses derived from the synthetic observations to the actual total dust masses present in the disk.  \newline
In \autoref{sec:method}, we describe our methods. The results are presented in \autoref{sec:results}. In \autoref{sec:discussion} we include a short discussion before we conclude the paper in \autoref{sec:conclusions}.

\section{Methods} 
\label{sec:method}
We carry out three-dimensional thermal two-fluid (gas+dust) hydrodynamic simulations of circumstellar disks with an embedded planet. We use the grid-based code {\scriptsize JUPITER} \citep{Szulagyi2016} and implement a pressureless solver to solve the dynamics of a dust fluid (assumed mm-sized grains) in addition to the gas fluid. We then processed the hydrodynamic simulation outputs with {\scriptsize RADMC-3D} \citep{Dullemond2012}, a wavelength-dependent radiative transfer tool to obtain intensity images on a given wavelength. In a second step, we used the Common Astronomy Software Applications package ({\scriptsize CASA}) \citep{CASA2007} to create the final synthetic mm-continuum images of the circumstellar disks with an embedded giant planet for the Atacama Large Millimeter/submillimeter Array (ALMA). In section \ref{sec:physmodel} we present the physical models of the gas and dust components and introduce the interaction terms which also include the back-reaction from dust onto the gas. We also introduce the treatment of radiation and cooling/heating mechanisms for the two fluids. In section \ref{sec:numericalmethod} we describe the numerical methods used to solve the hydrodynamic equations introduced in section \ref{sec:physmodel}. The details of our sets of thermal hydrodynamic simulations are presented in section \ref{sec:hydrodimulationssetup}. To conclude the method section, we present the two post-processing steps in section \ref{sec:post-processing}.

\subsection{Physical model}
\label{sec:physmodel}
\subsubsection{Gas and radiation}
We use the {radiative hydrodynamics} code {\scriptsize JUPITER} as presented in \cite{Szulagyi2016} to solve the hydrodynamic equations of the gas and radiation components in three dimensions on a spherical grid. In addition to the mass, momentum and energy equations (see equations (1) to (3) in \citealp{Szulagyi2016}), we describe the gas with an equation of state of an ideal gas. It relates the gas pressure $P_g$ to the internal energy {density} of the gas $e_g$ as
\begin{equation}
\label{eq:idealgaseos}
P_g=(\gamma-1)e_g
\end{equation}
with $\gamma=1.43$ being the adiabatic index. The energy equation describes the time evolution of the total energy of the gas per unit volume $E_g$ as the sum of radiation energy per unit volume $e_{\mathrm{rad}}$, the internal energy per unit volume $e_g$ and the kinetic energy per unit volume of the gas
\begin{equation}
    E_g=e_{\mathrm{rad}}+e_g+\frac{1}{2}\rho_g\mathbf{v}_g^2
\end{equation}
where $\rho_g$ is the density and $\mathbf{v}_g$ is the three-dimensional velocity vector of the gas. The fourth equation which governs the dynamics is the radiation equation (see equation (4) in \citealp{Szulagyi2016}). It describes the dynamics of the radiation energy $e_{\mathrm{rad}}$ and contains the flux-limited diffusion approximation with the two-temperature approach \citep[e.g.][]{Commercon2011}. \newline
The central star is assumed to be solar-like with radius $R_\star=R_\odot$, mass $M_\star=M_\odot$ and surface temperature $T =5780$ K.

\subsubsection{Dust}
In this work, we model a single dust size species of size $a$ as an additional pressureless fluid which has its distinct dynamics. Due to the ALMA continuum images we wanted to create, we choose the grain size to be 1 mm. The equations to describe a pressureless dust fluid are the Euler equations in the limit of vanishing sound speed ($c_s \to 0$) {\citep[e.g.][]{Paardekooper2006,Youdin2007}}. Thus, the system of mass and momentum equations is closed without an equation of state. The system of two equations in three dimensions, including source terms, has the following coordinate free form:
\begin{equation}\label{eq:dustmassconservation}
\frac{\partial \rho_d}{\partial t}+\bm{\nabla}\cdot(\rho_d \mathbf{v}_d)=0
\end{equation}
\begin{equation}\label{eq:dustmomentumconservation}
\frac{\partial (\rho_d\mathbf{v}_d)}{\partial t}+\bm{\nabla}\cdot(\rho_d\mathbf{v}_d\otimes\mathbf{v}_d)=-\rho_d\nabla\Phi+\mathbf{F}^{\mathrm{drag}}_d
\end{equation}
Here, $\rho_d$ is the dust volume density and $\mathbf{v}_d$ is the dust velocity. As opposed to the gas, the dust momentum equation~(\ref{eq:dustmomentumconservation}) does not contain a pressure term. We also do not include turbulent diffusion in the dust. The source terms on the right-hand side of the momentum equation~(\ref{eq:dustmomentumconservation}) contain the gradient of the gravitational potential $\Phi$. The gravitational potential contains contributions from the star, the planet {and the indirect term due to the motion of the star around the barycenter}. Besides the gravitational force, dust particles (and the gas) are also influenced by an aerodynamic drag term $\mathbf{F}^{\mathrm{drag}}_d$. Dust particles that are mixed in gas constantly collide with gas molecules. These collisions transfer momentum, influencing the dynamics of both the dust and the gas. We describe the degree of coupling between dust and gas with the dimensionless Stokes number $St$. We assume Epstein drag which is valid as long as the dust particles are smaller than the mean free path of the gas ($a<\lambda_\mathrm{mfp}$) which is valid for the range of disk parameters and dust particle size assumed in this study. The Stokes number of a dust particle of size $a$ in a gas with adiabatic index $\gamma$ in the Epstein regime takes the following form \citep[e.g][]{Laibe2012}: 
\begin{equation}
\label{eq:Stokesnumber}
St=\sqrt{\frac{\pi\gamma}{8}}\cdot\frac{\Omega_K\rho_{\bullet} a}{\rho_g c_s}
\end{equation}
Here, the Stokes number is a function of the local Keplerian angular velocity $\Omega_K$, the dust particle solid density $\rho_{\bullet}$ and the thermal sound speed in gas $c_s=\sqrt{\gamma k_B T/m_{\mu}}$ where $m_\mu=3.9\times10^{-24}$ g {is the mean mass of a gas molecule.} Hence, small dust particles, which are easily influenced by the gas have a small Stokes number. This is also true for dust particles in a high gas density environment. On the other hand, large dust particles or particles in a low-density environment, are influenced less by the gas and have a large Stokes number (as long as $St<1$). In this study, we only focus on 1 mm-sized dust particles. {In our models, these particles have Stokes number in the range $9\cdot10^{-3} <St <7\cdot10^{-2}$ in the disk midplane before inserting the planet. After the insertion of the planet, this range will become much broader due to the density fluctuation in gas. Moreover, in the disk regions above and below the midplane where the gas density drops off, the particles are less coupled and the Stokes numbers are generally larger than in the midplane.} For the grain composition, we assume a fractional abundance of 70\% silicate of solid density 3.5 g/cm$^3$ and 30\% refractory carbon of solid density 1.8 g/cm$^3$ \citep{Zubko1996,Li1997} which results in a solid density of the dust grains of $\rho_{\bullet}$ = 3 g/cm$^3$. \newline 
The Stokes number, as defined in equation~(\ref{eq:Stokesnumber}), is used to parametrize the drag force $\mathbf{F}^{\mathrm{drag}}$ which is responsible for the exchange of momentum between gas and dust \citep{Weidenschilling1977}. The drag contributions to the gas and the dust fluid are symmetric and also depend on the relative velocity between the two fluids $\mathbf{v}_d-\mathbf{v}_g$ as: 
\begin{equation} 
\label{eq:modeldustdragforce}
\mathbf{F}^{\mathrm{drag}}_{d}=-\mathbf{F}^{\mathrm{drag}}_{g}=-\rho_d\frac{\Omega_K}{St}(\mathbf{v}_d-\mathbf{v}_g)
\end{equation}
Here, $\mathbf{F}^{\mathrm{drag}}_{d}$ is the drag force acting on the dust caused by the movement through the gas and $\mathbf{F}^{\mathrm{drag}}_{g}$ is the drag force acting on the gas caused by the movement through dust \citep[e.g.][]{Whipple1972}. 

\subsubsection{Planet}
The embedded planet is modeled solely via its gravitational potential as a point mass:
\begin{equation}
    U_p(x,y,z) = \frac{G M_p}{\sqrt{(x-x_p)^2+(y-y_p)^2+(z-z_p)^2+r_s^2}}
\end{equation}
where $G$ is the gravitational constant, $M_p$ is the mass of the planet, $(x_p,y_p,z_p)$ are the coordinates of the planet in Cartesian coordinates. $r_s$ is a smoothing length to avoid singularities in the potential at the location of the planet. We set the smoothing length to the length of one cell diagonal of the computational grid (see \autoref{sec:hydrodimulationssetup}). 

\subsection{Numerical method}
\label{sec:numericalmethod}
The main computational tool which we use to solve the hydrodynamic equations is the {\scriptsize JUPITER} code. It was originally developed by F. Masset and J. Szul\'{a}gyi. For this study, we added a numerical dust solver to solve for a second (pressureless) fluid (dust) and its interaction with the gas. The {\scriptsize JUPITER} code is a three-dimensional Godunov-type code that solves the hydrodynamic equations on a grid using Riemann solvers. Even though the {\scriptsize JUPITER} code has nested mesh capability, it is not used in this study, so the planet vicinity is unresolved. The radiative module of the code applies a flux-limited diffusion approximation with the two-temperature approach as described in \cite{Szulagyi2016} and  \cite{Szulagyi2018}. We include thermal processes in the gas such as adiabatic heating/cooling, viscous heating and radiative cooling, as well as stellar irradiation. For both, the gas and the dust, we use an operator splitting method to solve advection terms separately from the source terms. When solving the pressureless equations, we apply the method described by \cite{LeVeque2004} which was also used by \cite{Paardekooper2006} to run two-fluid (gas+dust) simulations of circumstellar disks. As in other Riemann solvers \citep{Toro2009}, the analytic solution to the Riemann problem of the system of equation~(\ref{eq:dustmassconservation}) and equation~(\ref{eq:dustmomentumconservation}) is the basis of the numerical method. The solution to the Riemann problem for a pressureless fluid is significantly different from the solution in gas which includes pressure \citep{Bouchut2003,LeVeque2004}. It consists of a single wave moving from the Riemann interface at speed
\begin{equation}
    \hat{u} = \frac{\sqrt{\rho_L}u_L+\sqrt{\rho_R}u_R}{\sqrt{\rho_L}+\sqrt{\rho_R}}
\end{equation}
where $\rho_L$ and $u_L$ are the density and velocity on the left-hand side of the Riemann interface and $\rho_R$ and $u_R$ are the density and velocity on the right-hand side of the interface. The Riemann fluxes are determined based on the sign of $\hat{u}$ at every interface according to the following scheme: 
\begin{equation}
\label{eq:numericalinterfaceflux}
\mathbf{F}_{\text{int}}=
    \begin{cases}
        \mathbf{F}_L & \text{ if}\;\hat{u}>0\\
        \frac{1}{2} \big(\mathbf{F}_L +\mathbf{F}_R \big) &\text{ if}\; \hat{u}=0\\
         \mathbf{F}_R &\text{ if}\; \hat{u}<0
    \end{cases}
\end{equation}
Following \cite{LeVeque2004}, we add correction terms to the interface flux to achieve second-order accuracy on smooth solutions and apply the \textit{minmod} flux-limiter to avoid spurious oscillations around discontinuities \citep{LeVeque2010}. \newline
In the numerical source step, we deal with the interaction between dust and gas. Following the operator splitting scheme, the two equations which we solve are: 
\begin{equation} 
\frac{\partial \mathbf{v}_g}{\partial t}=\frac{\mathbf{F}^{\mathrm{drag}}_{g}}{\rho_g}
\label{eq:source_eq1}
\end{equation}
and
\begin{equation} 
\frac{\partial \mathbf{v}_d}{\partial t}=\frac{\mathbf{F}^{\mathrm{drag}}_{d}}{\rho_d}
\label{eq:source_eq2}
\end{equation}
with definitions of the drag force $\mathbf{F}^{\mathrm{drag}}$ as in equation~(\ref{eq:modeldustdragforce}). We find the solutions to equations~(\ref{eq:source_eq1}) and (\ref{eq:source_eq2}) using an implicit finite difference scheme as described in \cite{Benitez-Llambay2018}. The scheme leads to an update formula for the velocities $\mathbf{v}_g$ and $\mathbf{v}_d$ as in \cite{Stone1997} equations (6) and (7). The velocities are updated in the source step along with the other source terms (gravitational and fictitious force terms). {The implementation of the dust solver was tested\footnote{Tests were performed within the framework of a master's thesis at ETH Zürich. The thesis can be found here: \url{people.phys.ethz.ch/~judits/Ms_Binkert.pdf}} against analytic Riemann solutions as suggested in \citep{LeVeque2004} and compared to previous two-fluid studies \citep[e.g.][]{Paardekooper2006}}. Further, we modified the opacity $\kappa(T,\rho_g)$ used in \cite{Szulagyi2016}. In the one-fluid code, the opacity is calculated based on the local gas density $\rho_g$ assuming a dust-gas mixture with a locally constant dust-to-gas ratio $dtg=0.01$. In the two-fluid simulations, we reduce the opacity where the dust-to-gas ratio is smaller and {increase} the opacity where the dust-to-gas ratio is larger. We computed the local two-fluid opacity $\kappa_\textrm{2f}$ based on the local dust density $\rho_d$ and gas density $\rho_g$ as

\begin{equation}
    \kappa_\textrm{2f}(T,\rho_g,\rho_d) = \kappa\Big(T,0.01\rho_g+0.99\frac{\rho_d}{dtg}\Big)
\end{equation}
where $dtg=0.01$ is the dust-to-gas ratio assumed in the one-fluid dust-gas-mixture. 
\subsection{Hydrodynamic simulation setup}
\label{sec:hydrodimulationssetup}
\subsubsection{Disk setup and simulation domain}
We set up a disk with a gas surface density following a power law as
\begin{equation}
    \Sigma_g(r)=\Sigma_{g,0}\cdot\bigg( \frac{r}{\text{AU}} \bigg)^{-1/2}
\end{equation}
where $\Sigma_{g,0}$ = 80g/cm$^2$ is the gas surface density at 1 AU. This disk contains a total gas mass of $\sim$ 0.05 M$_\odot$ ($\sim$52 M$_\textrm{jup}$) between 1 AU and 120 AU. With this disk setup, the Toomre Q parameter, which is a criterion for disk instability \citep{Toomre1964}, remains well above Q $>$ 1.7 in all our simulations. Hence, it can be expected that the disk self-gravity is negligible. \cite{Humphries2018,Humphries2019} have preformed three-dimensional global (SPH) disk simulations including a planet in situations when gravitational instability is important. 
\newline 
We use a constant kinematic viscosity $\nu$ = 3.15$\cdot$10$^{15}$cm$^2$/s. Assuming an isothermal disk with aspect ratio H = 0.05, this is equivalent to a Shakura \& Sunyaev $\alpha$-parameter of $\alpha$ = 4.0$\cdot$10$^{-3}$ at 50 AU from a solar-mass star or $\alpha$ = 5.2$\cdot$10$^{-3}$ at 30 AU or $\alpha$ = 1.2$\cdot$10$^{-2}$ at 5.2 AU respectively \citep{Shakura1973}. We note here that our disk is not isothermal and that the aspect ratio varies depending on the local heating and cooling conditions in a range between 0.025 and 0.05. \newline
In addition to the gas fluid, we initialize a dust fluid that represents a single dust size species of 1 mm-sized dust particles. Initially, the dust-to-gas mass ratio is 0.01 everywhere, but during the disk evolution, this changes from location to location. \newline
The computational grid is set up identically to \cite{Szulagyi2016}, meaning we solve the hydrodynamic equations on a spherical grid (r, $\phi$ ,$\theta$) centered on the star. The frame of reference is co-rotating with the planet which orbits at distance $r_p$ with Keplerian angular frequency $\Omega_K = \big(G(M_*+M_p)/r_p^3\big)^{1/2}$ where $M_* = M_\odot$ is the mass of the central star. Hence, the planet always remains {fixed on the} grid. To save computational costs, we do not simulate the entire disk between 1 AU and 120 AU in every simulation, but only the range between 0.4$r_p$ to 2.4$r_p$. In the case of a planet orbiting at $r_p$ = 50 AU this corresponds to a range between 20 AU and 120 AU. In azimuthal direction, the simulation covers the full range from $\phi_{min}=-\pi$ to $\phi_{max}=\pi$. The opening angle of the grid is set to $\theta_0=7.4^\circ$. We assume the disk to be symmetric about the midplane and constrain the simulation to polar angles between {$\theta_{min}=\pi/2-\theta_0$ and $\theta_{max}=\pi/2$} to further save computational time. Our grid consists of $N_r=215$ radial, $N_{\phi}=680$ azimuthal and $N_{\theta}=20$ polar cells {, which creates roughly cubical grid-cells. A further increase in resolution would increase the computational cost for this study by an unreasonable amount.}

\subsubsection{Initial/boundary conditions and simulation procedure}
\label{sec:procedure}
We initialize the gas disk with a constant aspect ratio $H = h_g/r = 0.05$ and the 1-mm sized dust {such that the dust-to-gas ratio is uniformly at 0.01}. This all of course evolves during the simulation, as the disk evolves. Before we introduce the planet to the disk, we evolve the disk without the planet and only 2 cells in azimuthal direction (since the disk without planet is azimuthally symmetric) for 150 planetary orbits. This allows the system to reach thermal equilibrium and the 1 mm-sized dust to settle vertically to reach a quasi-steady state. Then, we divide the 2 azimuthal cells into 680 cells and introduce the planet. We increase the mass of the planet over the following 100 orbits until it reaches its final mass to not introduce unwanted perturbations. We then evolve the system for another 100 orbits to arrive at a total of 200 planetary orbits. The boundary conditions {for the gas} are identical to \cite{Szulagyi2016}.
{In detail, at the radial boundaries, the density and energy were extrapolated based on the initial slope and the value in the adjacent active cell. At the radial boundaries, the radial and polar velocity component in the ghost cells were set equal to the value in the adjacent active cell, i.e. symmetric boundary conditions. The azimuthal velocity component was extrapolated based on the local Keplerian velocity and the value in the adjacent active cell.} In polar direction, we used reflective, i.e. anti-symmetric, boundary conditions. At the upper polar boundary, the temperature {in the ghost cells} was fixed at {3K} which accounts for the radiative cooling of the disk to outer space. We used periodic boundary conditions in azimuthal direction.
For the dust fluid, at the radial boundaries, we also used anti-symmetric boundary conditions for the radial velocity component. This prevents inflow and outflow of dust in the simulation domain. The density and the other velocity components have symmetric boundary conditions at the radial boundaries. The boundaries condition in azimuthal and polar direction were set equal to the boundary conditions in gas, except for the dust density at the polar boundary opposite to the midplane. There, the dust density was set to a floor value. The floor value corresponds to about one mm-sized dust grain per computational cell.

\subsubsection{Simulation sets}
\label{sec:hydrosimulations}
We carry out a set of twelve radiative hydrodynamic simulations which are summarized in \autoref{table:simulationsets}. We choose four different planetary masses (5 M$_{\text{jup}}$, 1 M$_{\text{jup}}$, 0.3 M$_{\text{jup}}$, 0.05 M$_{\text{jup}}$, the latter two are equivalent to the mass of Saturn and the mass of Neptune respectively) which we place at three different radii (5.2 AU, 30 AU and 50 AU). After injecting a planet, we evolve each simulation for 200 planetary orbits as described in section \ref{sec:procedure}. 

\begin{table}
\begin{center}
\begin{tabular}{l c c c} 
 \hline
  simulation & $r_p$ & $M_p$  & $\Sigma_{g,t=0}(r=r_p)$   \\ 
  &(AU) & (M$_{\text{jup}}$)& (g/cm$^2$)\\
  \hline
 \texttt{m5au1nep} &  5.2  & 0.05 & 35 \\ 
 \texttt{m5au1sat} &  5.2  & 0.3  & 35 \\ 
 \texttt{m5au1jup} &  5.2  & 1  & 35 \\ 
 \texttt{m5au5jup} &  5.2  & 5  & 35 \\ 
 \texttt{m30au1nep} &  30  & 0.05  & 15 \\ 
 \texttt{m30au1sat} &  30  & 0.3  & 15 \\ 
 \texttt{m30au1jup} &  30  & 1.0  & 15 \\ 
 \texttt{m30au5jup} &  30  & 5.0  & 15 \\ 
 \texttt{m50au1nep} &  50 & 0.05  & 11 \\
 \texttt{m50au1sat} &  50  & 0.3  & 11 \\
 \texttt{m50au1jup} &  50  & 1.0  & 11 \\
 \texttt{m50au5jup} &  50  & 5.0  & 11 \\
 \hline
\end{tabular}
\caption{This table provides an overview of the 12 hydrodynamic simulations which we carried out and the parameters used. We varied the planetary orbital radius $r_p$ between 5.2 AU and 50 AU and the planetary mass $M_p$ between Neptune mass (0.05 M$_\mathrm{jup}$) and 5 Jupiter masses as shown in the second and third column. The last column shows the initial gas surface density $\Sigma_{g,t=0}$ at the location of the planet.}
\label{table:simulationsets}
\end{center}
\end{table}

\subsection{Post-processing}
\label{sec:post-processing}
For each of the hydrodynamic simulations, we created synthetic ALMA mm-continuum observations. As in \cite{Szulagyi2018}, we process our models with {\scriptsize RADMC-3D} (v0.41), a radiative transfer tool developed by \cite{Dullemond2012} and the Common Astronomy Software Applications package\footnote{\url{casa.nrao.edu}} ({\scriptsize CASA}). In a first step, we compute the dust temperature with a thermal Monte Carlo approach using {\scriptsize RADMC-3D} which assumes that the dust is in radiative equilibrium with the radiation field. Then, we perform ray-tracing with {\scriptsize RADMC-3D} to generate intensity images of our disk models at different wavelengths and create synthetic ALMA images of the disks using {\scriptsize CASA}. 

\subsubsection{RADMC-3D}
Not only the density distribution in the disk has a large impact on the mm-continuum observations but also the temperature structure. We determine the dust temperature in the disk using the \texttt{mctherm} task of {\scriptsize RADMC-3D} which performs a thermal Monte Carlo simulation. As in the hydrodynamic simulations, we assume the radiation source to be a solar-like star of mass 1~M$_{\sun}$, radius 1~R$_{\sun}$ and temperature T$_{\text{eff}}$~=~5780K. The radiation field of the star is represented by $2.1\cdot10^9$ photon packages which are emitted isotropically before they travel through the disk and are scattered, absorbed and re-emitted along their path by dust grains until they eventually leave the model. The temperature computed in this process is the equilibrium temperature of the dust in the radiation field of the central star. In our setup, the dust does not acquire thermal energy from the gas. \newline
Even though the hydrodynamic simulations are radiative, the radiation is treated in a wavelength-independent way. To obtain wavelength-dependent intensity images of our disk models, we apply the {\scriptsize RADMC-3D} \texttt{image} task {for which we also set scattering to be isotropic}. The Python pipeline used to convert the {\scriptsize JUPITER} code output files is based on \cite{Szulagyi2018}. We set up the radiative transfer with the same stellar source as in the thermal Monte Carlo simulation to be consistent throughout our hydrodynamic simulations and post-processing steps. For the dust temperature, we use the temperature determined in the previous thermal Monte Carlo simulation which is not necessarily equal to the gas temperature. The opacity table provided to {\scriptsize RADMC-3D} is identical to the one used in \cite{Szulagyi2018} and is based on a dust mixture of 70\% silicate and 30\% carbon. It was computed considering Mie theory using the BHMIE code of \cite{Bohren1984} assuming a dust grain size distribution of 0.1 $\micron$ and 1 cm with a power-law index of 3.5. The size of the intensity image is set to 1000 $\times$ 1000 pixels and the distance between the observer and the disk is assumed to be 100 parsec which is similar to the distance to the closest star-forming regions.

\subsubsection{Synthetic ALMA observations}
We process the intensity images as generated by{\scriptsize RADMC-3D} with {\scriptsize CASA} and create synthetic ALMA observations using the \verb|simobserve| and \verb|simanalyze| tasks. Furthermore, we use ALMA cycle 7 array configurations which have baselines ranging from 0.16 km to 16.2 km and allow us to explore different beam sizes. For each antennae configuration, we create observations at different wavelengths bands to explore the optimal observing setups. The channel bandwidth of our continuum observations is 7.4GHz. The integration time is chosen to be 300s per pointing with a total integration time of 3h. We add thermal noise to the synthesized images using the tsys-atm parameter which constructs an atmospheric profile at the ALMA site at an altitude of 5000m, atmospheric pressure of 650mBar and 20\% relative humidity. The precipitable water vapor is set to 0.475mm and the ambient temperature is 269K.

\begin{figure*}
\includegraphics[width=1.5\columnwidth]{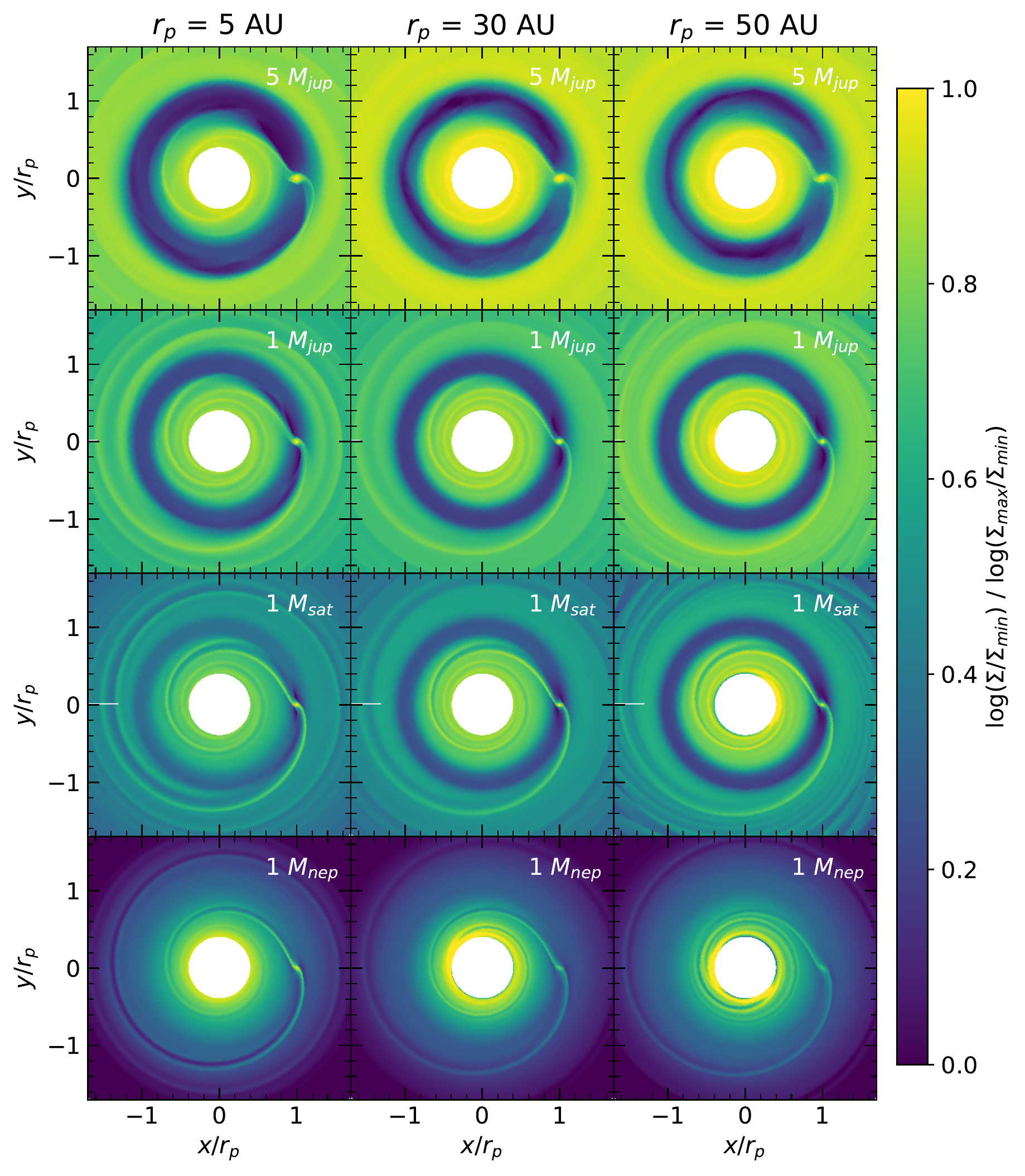}
\caption{The figure shows the normalized gas surface density distribution in logarithmic scale of our 12 hydrodynamic two-fluid simulations of a circumstellar disk with an embedded planet at 200 planetary orbits in a face-on view. In each panel, the columns show simulations with planets orbiting at different radii, $r_p$ = 5 AU, 30 AU, 50 AU. Each row shows the disk containing a planet with different mass ($M_p$ = 5 $M_{\text{jup}}$, 1 $M_{\text{jup}}$, 1 $M_{\text{sat}}$, 1 $M_{\text{nep}}$). The mass of the planet is indicated in the upper right corner of each subplot. The x and y coordinates are normalized with the planetary orbital radius $r_p$.}
\label{fig:surfacedensity_overview_gas}
\end{figure*}

\begin{figure*}
\includegraphics[width=1.5\columnwidth]{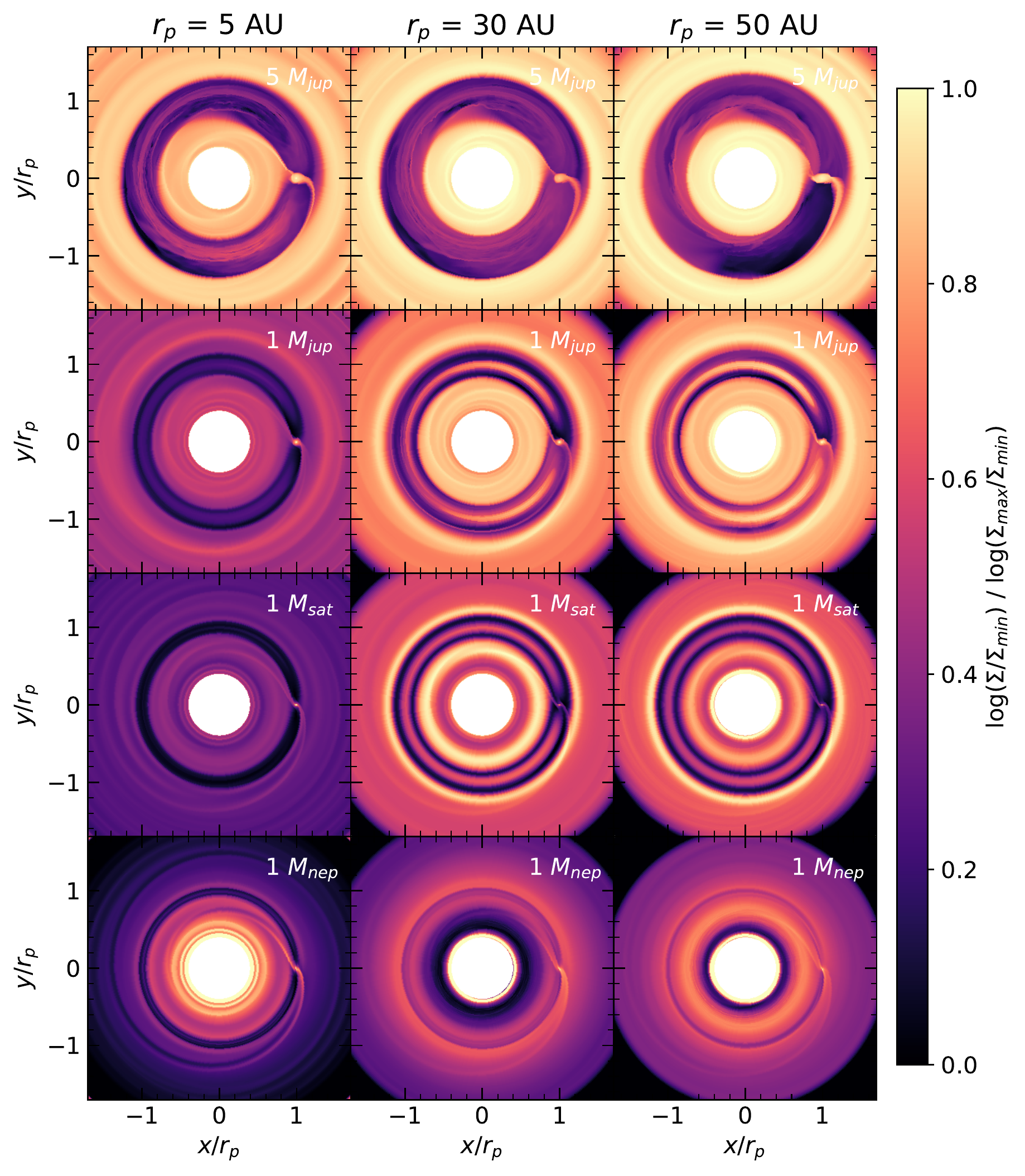}
\caption{The figure shows the normalized surface density distribution of 1 mm-sized dust in logarithmic scale of our 12 hydrodynamic two-fluid (gas+dust) simulations of a circumstellar disk with an embedded planet at 200 planetary orbits in a face-on view. In each panel, the columns show simulations with planets orbiting at different radii, $r_p$ = 5 AU, 30 AU, 50 AU. Each row shows the disk containing a planet with different mass ($M_p$ = 5 $M_{\text{jup}}$, 1 $M_{\text{jup}}$, 1 $M_{\text{sat}}$, 1 $M_{\text{nep}}$). The mass of the planet is indicated in the upper right corner of each subplot. In the x and y coordinates are normalized with the planetary orbital radius $r_p$.}
\label{fig:surfacedensity_overview_dust}
\end{figure*}

\begin{figure*}
\centering

\includegraphics[width=.7\textwidth]{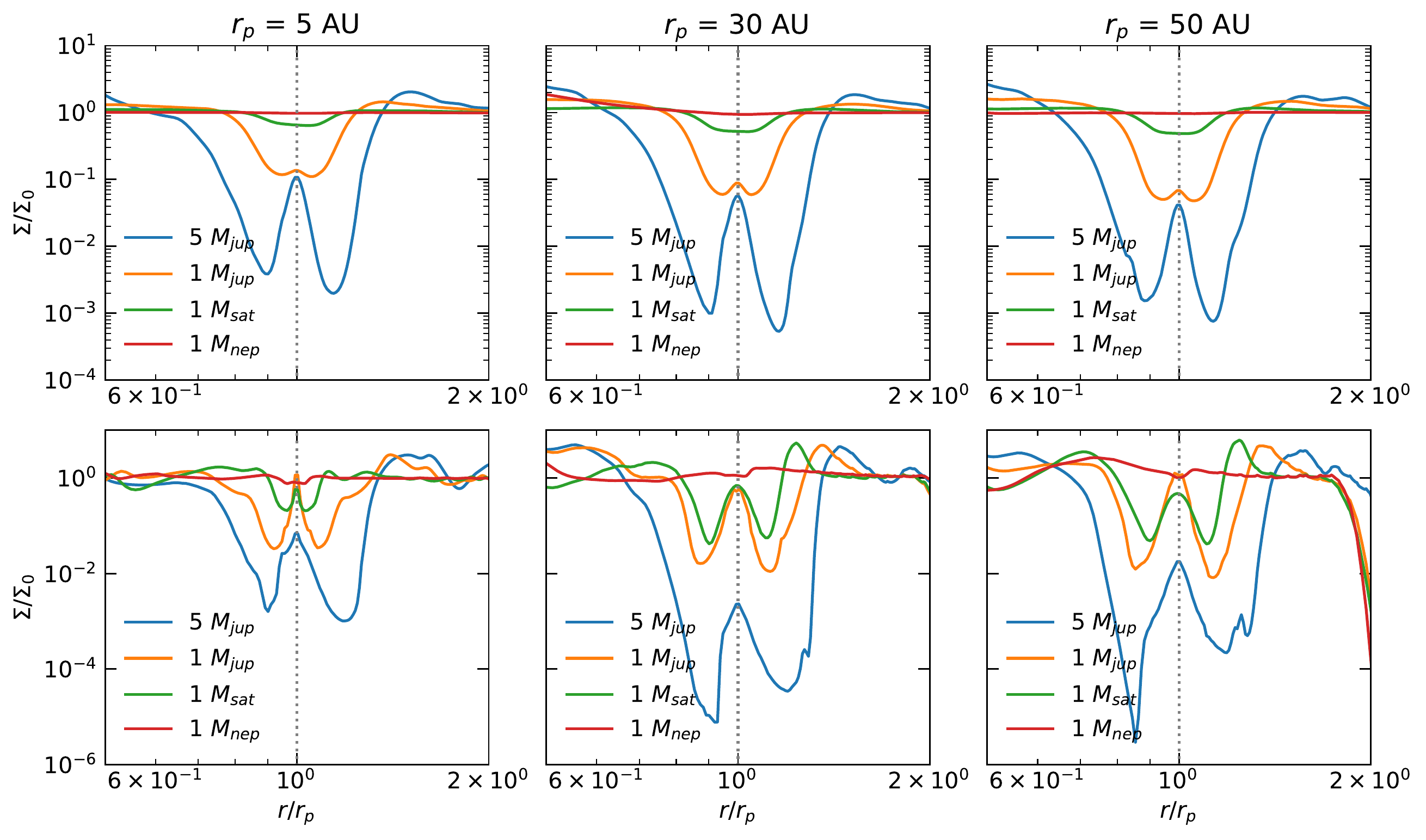}
\caption{The azimuthally averaged surface density in gas (top row) and 1 mm-sized dust (bottom row) of our 12 hydrodynamic simulations are shown here. Each panel shows profiles of 4 identical disks, each containing a planet of different mass ($M_p$ = 5 $M_{\text{jup}}$, 1 $M_{\text{jup}}$, 1 $M_{\text{sat}}$, 1 $M_{\text{nep}}$).  From left to right, the panels show planets orbiting at different radii, $r_p$ = 5 AU, 30 AU, 50 AU. The location of the planet is indicated with a dashed vertical line.}
\label{fig:surfdensprofiles_overview}
\end{figure*}

\begin{figure} 
\includegraphics[width=\columnwidth]{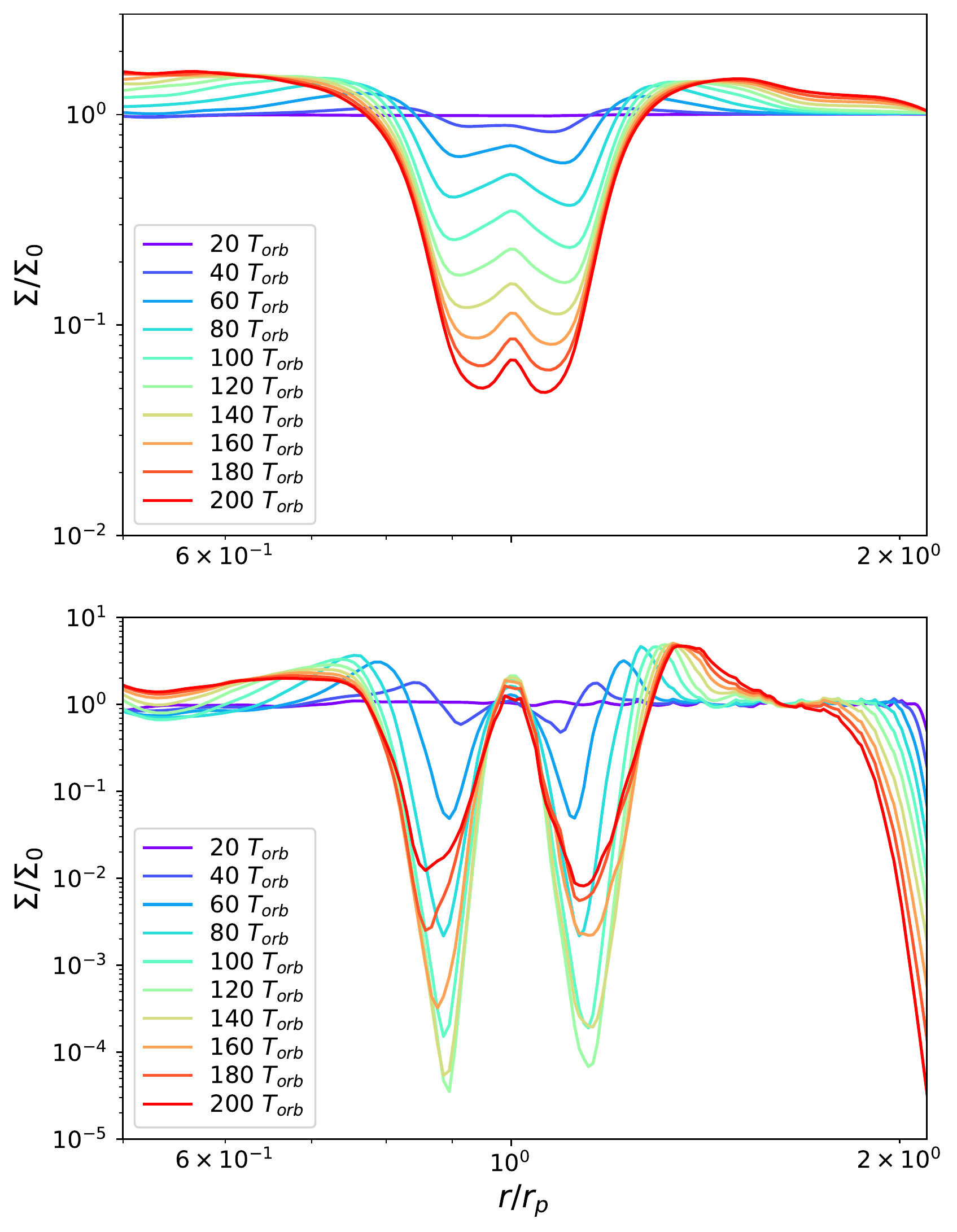}
\caption{Temporal evolution of the azimuthally averaged surface density profile in gas (top) and dust (bottom) in our \texttt{m50au1up} simulation containing a 1 Jupiter-mass planet orbiting at $r_p$ = 50 AU over the period of 200 planetary orbits. Plotted are the surface density profiles after every 20 orbits.}
\label{fig:rainbowplot}
\end{figure}

\begin{figure*}
\includegraphics[width=1.5\columnwidth]{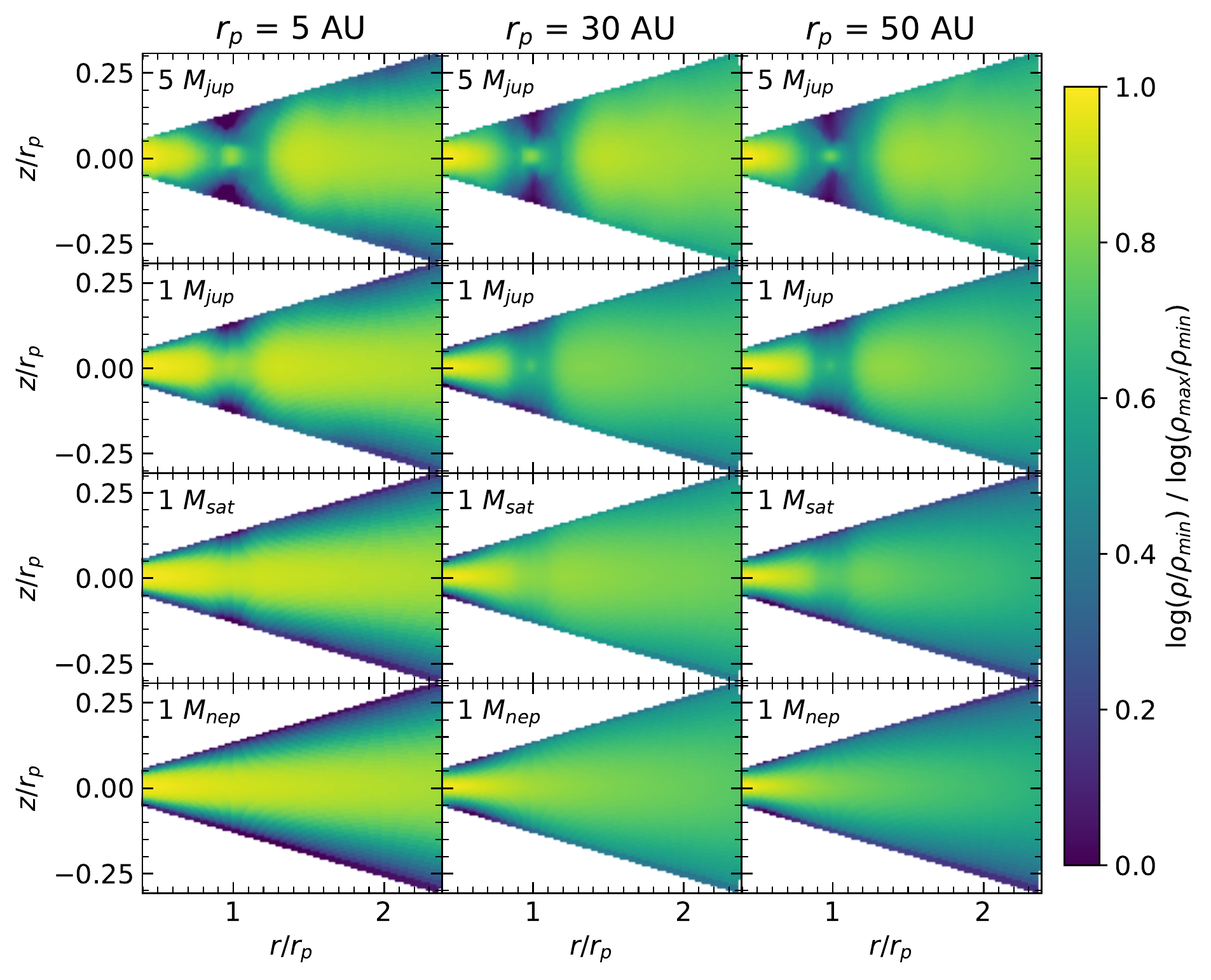}
\caption{Here we show the normalized vertical distribution of the azimuthally averaged gas volume density of our 12 hydrodynamic two-fluid simulations of a circumstellar disk with an embedded planet at 200 planetary orbits in a side-on view. The opening angle of the disks are enlarged for better visualization, i.e. the aspect ratio of the disk plotted larger. In the left panel we show the surface density in gas, in the right panel we show the surface density in 1mm-sized dust. In each panel, the columns show simulations with planets orbiting at different radii, $r_p$ = 5 AU, 30 AU, 50 AU. Each row shows the disk containing a planet with different mass ($M_p$ = 5 $M_{\text{jup}}$, 1 $M_{\text{jup}}$, 1 $M_{\text{sat}}$, 1 $M_{\text{nep}}$). The mass of the planet is indicated in the upper left corner of each subplot. }
\label{fig:verticaldistr_overview_gas}
\end{figure*}

\begin{figure*}
\includegraphics[width=1.5\columnwidth]{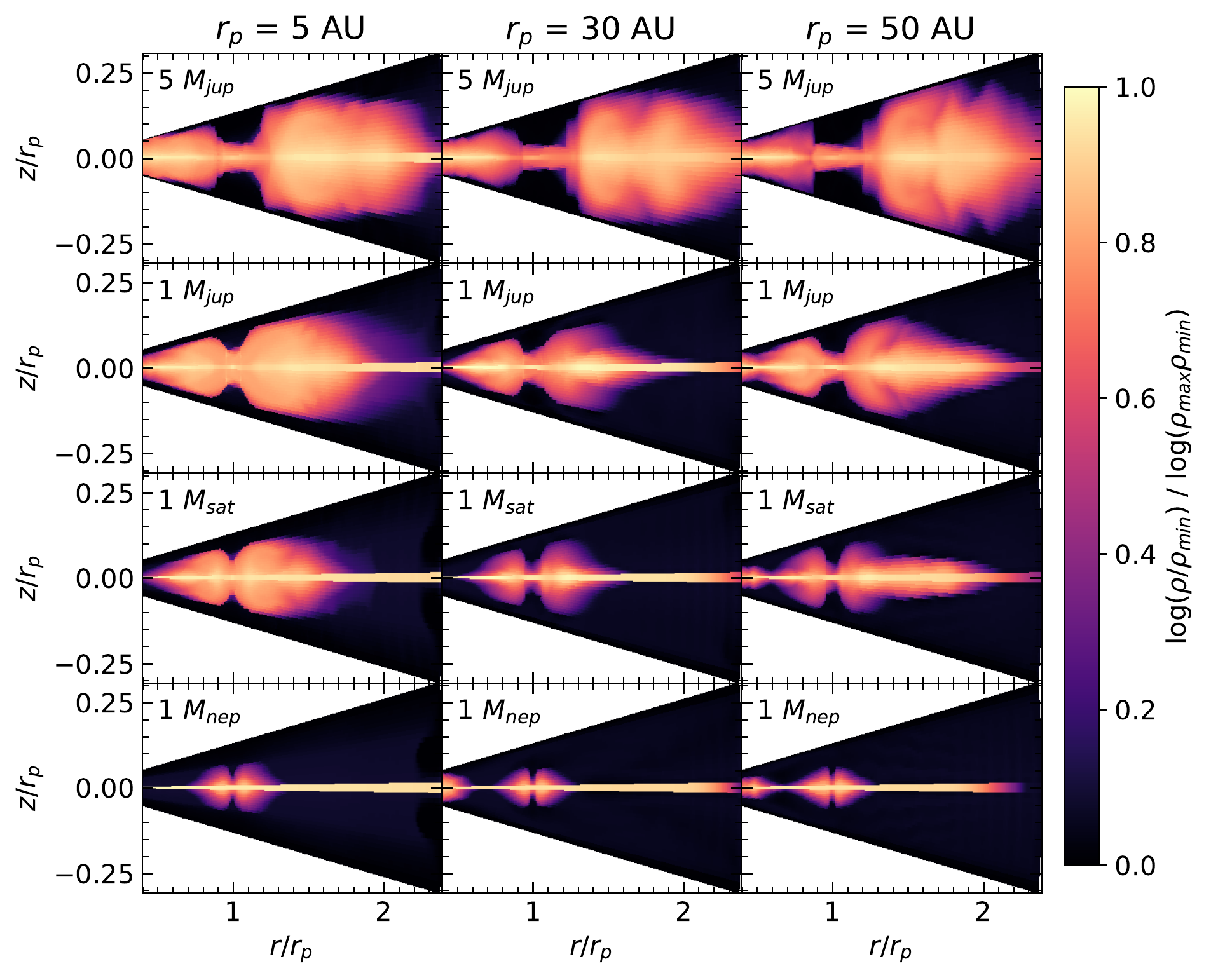}

\caption{Here we show the normalized vertical distribution of the azimuthally averaged volume density of our 12 hydrodynamic two-fluid simulations of a circumstellar disk with an embedded planet at 200 planetary orbits in a side-on view. The opening angle of the disks are enlarged for better visualization, i.e. the aspect ratio of the disk plotted larger. In the left panel we show the surface density in gas, in the right panel we show the surface density in 1 mm-sized dust. In each panel, the columns show simulations with planets orbiting at different radii, $r_p$ = 5 AU, 30 AU, 50 AU. Each row shows the disk containing a planet with different mass ($M_p$ = 5 $M_{\text{jup}}$, 1 $M_{\text{jup}}$, 1 $M_{\text{sat}}$, 1 $M_{\text{nep}}$). The mass of the planet is indicated in the upper left corner of each subplot. }
\label{fig:verticaldistr_overview_dust}
\end{figure*}


\begin{figure}
\includegraphics[width=0.8\columnwidth]{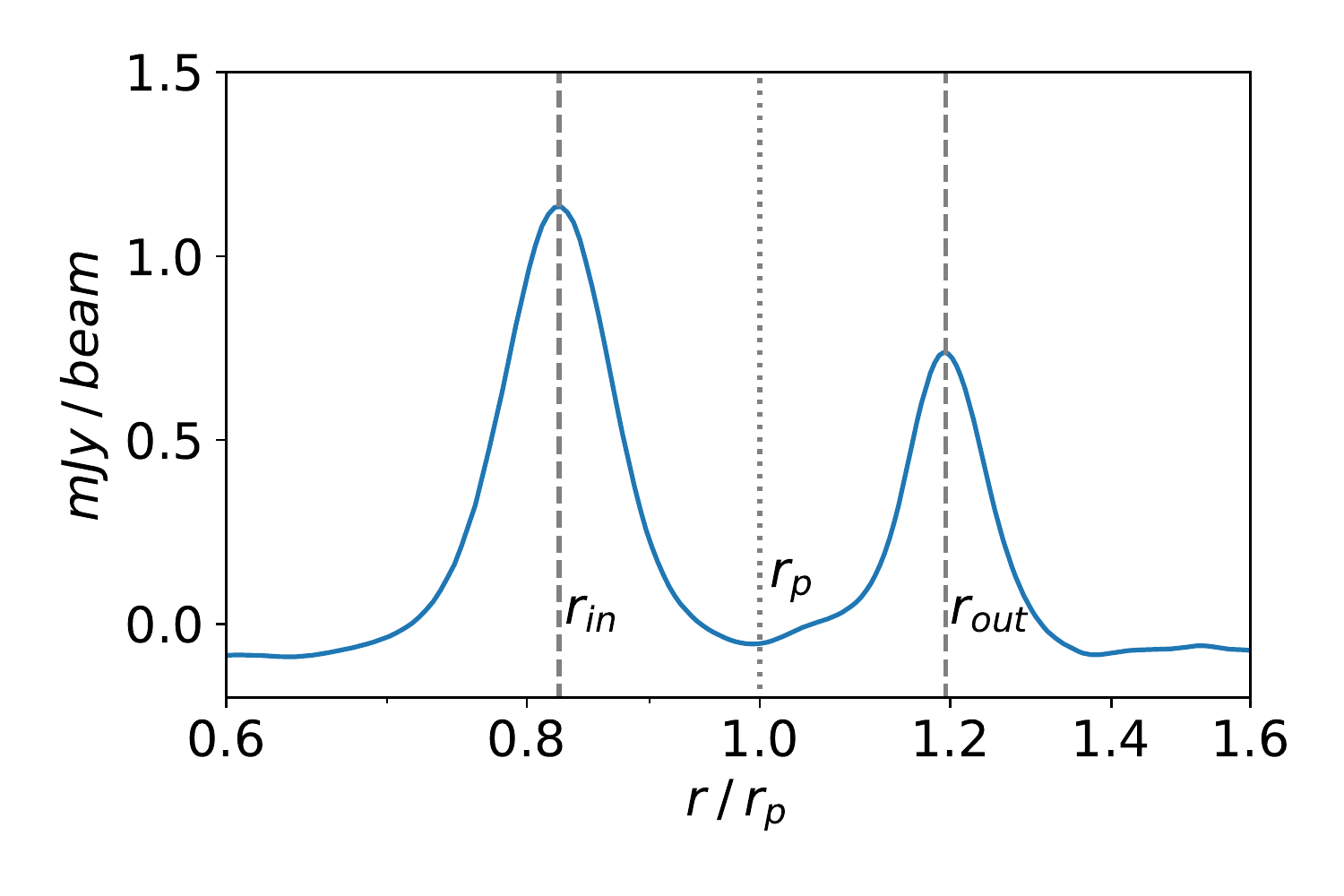}
\caption{Example of the gap width $\Delta$ as defined in equation (\ref{eq:gap_width_definition}). Shown is the azimuthally averaged intensity profile computed using the {\scriptsize CASA} output. We identify $r_{\mathrm{out}}$ and $r_{\mathrm{in}}$ as the location of the local maxima in the intensity profile. This particular figure shows the intensity profile of the simulation \texttt{m30au1sat} observed in ALMA band 9 (400-500\micron) with antennae configuration C43-8.}
\label{fig:aziavg_intensity}
\end{figure}

\section{Results}
\label{sec:results}
In the following section, we present the results of our three-dimensional two-fluid hydrodynamic simulations. This is followed by the results of the synthetic observations for ALMA. Analysis and discussion of the results will be presented in the next section. 

\subsection{Hydrodynamic simulations}
We run a set of twelve three-dimensional tow-fluid (gas + 1 mm-sized dust) hydrodynamic simulations of a circumstellar disk with an embedded planet. We place a planet of four different masses in an orbit at three different semi-major axes. In this section, we present the resulting effects the planets have on the gas and dust distribution on their hosting circumstellar disk. 

\subsubsection{Surface density}
\label{sec:result_gas}
In \autoref{fig:surfacedensity_overview_gas} we plot the vertically integrated volume density, i.e. the surface density, of the gas component at 200 planetary orbits. In \autoref{fig:surfacedensity_overview_dust} we show the corresponding surface density in 1 mm-sized dust. The figures show the surface density in Cartesian coordinates, even though the computations are carried out on a spherical grid, with the star located at the center of each sub-panel and the planet at the three o'clock position orbiting in a counterclockwise direction. In both, \autoref{fig:surfacedensity_overview_gas} and \autoref{fig:surfacedensity_overview_dust}, the columns show the simulations containing planets with different semi-major axis ($r_p$ = 5 AU, 30 AU, 50 AU). At these orbital radii, a period of 200 planetary orbits is equivalent to $\sim$ 2.4 kyr, $\sim$ 33 kyr or $\sim$ 71 kyr respectively. The four rows show the simulations containing a planet with mass $M_p$ = 5 $M_{\text{Jup}}$, 1 $M_{\text{Jup}}$, 1 $M_{\text{Sat}}$, 1 $M_{\text{Nep}}$. The color-map of the figures has a logarithmic scale and is normalized so that we can use the same color-map for all the sub-panels in gas and dust respectively.
In both, the gas and the dust disk, the planet creates distinctive disk morphologies. While for planets masses $M_p\geq1M_\text{sat}$, a gap is seen in the gas, a gap is seen in dust also when $M_p=1M_\text{nep}$. \newline
In \autoref{fig:surfdensprofiles_overview} we show the azimuthally averaged surface density profiles in gas (\emph{top row}) and dust (\emph{bottom row}). From left to right, we plot the surface density profiles with the planet at $r_p$~=~5.2~AU, 30 AU and 50 AU at 200 planetary orbits. \newline
Across our simulations, we vary the planet mass from 1 $M_\text{nep}\approx$~0.05~$M_\text{jup}$ to 5 $M_\text{jup}$. Similar to \cite{Paardekooper2006}, we find that the $M_p$ = 1 $M_\text{nep}$ planet does only open a gap in 1 mm sized dust but not in gas. The Neptune-mass planet is able to disturb the gas surface density by producing distinct spiral waves. However, the gravitational torque caused by the planet is not large enough to overcome the viscous effects in the gas. 
The gap which is opened by the 1 $M_\text{nep}$ in the dust is very shallow. Increasing the planet mass decreases the gas surface density in the vicinity of the planetary orbital radius because gravitational torques increase and push the gas away from the planetary orbit. As expected, the depth and width of the gap increase with the mass of the planet. We find that the depth of the dust gap is generally deeper than the depth in gas.\newline
At the outer edge of the dust gap, a density enhancement forms which coincides with a pressure maximum in gas. From both sides of the pressure maximum, dust drifts towards it. Hence, dust in the outer disk which drifts inwards gets trapped at this location, steadily increasing the local dust density. We show this effect in \autoref{fig:rainbowplot} where we plot the temporal evolution of the azimuthally averaged surface density distribution in gas (\emph{top}) and dust (\emph{bottom}) for every 20 orbits in the case of the 1 Jupiter-mass planet orbiting at 50 AU. The dust enhancement at the outer edge of the gap also becomes broader with time and moves outward following the gas pressure maximum. The dust enhancement is generally broader for more massive planets. The maximum dust density remains roughly constant for different planetary masses for $M_p\geq$ 1 $M_\text{sat}$. As shown in \autoref{fig:rainbowplot}, we find that {the width of the} gas gap quickly approaches a quasi-steady state. The depth of the gas gap increases monotonically over time and the width of the gap approaches its final value after about 100 orbits when the planet reaches its final mass. The gas disk still continues to evolve viscously and we expect a true steady-state to be reached in the gas after one viscous timescale $t_\mathrm{vis}=x^2/\nu$ \citep[e.g.][]{Lynden-Bell1974}. Over the length scale of a gas gap with width $\Delta_\mathrm{gas}=0.5\cdot r_p = 0.5\cdot x$, the viscous timescale in our simulations is $\sim 2\cdot 10^3$ orbital timescales at 5.2 AU. However, \cite{Kanagawa2017} find that the width of the gas gap does only change about 10\% after $0.1\cdot t_\mathrm{vis}$ which is equivalent to $\sim 200$ orbital timescales at 5.2 AU. On the other hand, the gap profile in dust evolves continuously and the features in the 1 mm-sized dust disk change over time. Interestingly, the depth of the dust gap in our \texttt{m50au1jup} simulation does reach a maximum already after about 120 orbits, after that time, the depth decreases again. We find this behavior in all our simulations with  $M_p \geq$ 1 $M_{\text{Jup}}$. In some cases, a pressure maximum also forms in the inner disk, e.g in the \texttt{m30au1sat} simulation. In that case, a dust density enhancement also forms in the inner disk because the 1 mm-sized dust drifts towards the pressure maximum at this location as well. \newline
Intermediate mass planets ($M_p$ = 1 $M_\text{sat}$, 1 $M_\text{jup}$) show the distinct W-shaped dust surface density profile as found in previous studies, e.g. by \cite{Dipierro2016}. There exists a substantial amount of dust in the co-orbital region of the planet which also causes a peak in the surface density profile in \autoref{fig:surfdensprofiles_overview} at the location of the planet. Since for these masses, a gap is also opened in the gas, drag is reduced due to the low gas density and there exists a region around the orbital radius of the planet {in which effects of radial drift are smaller and dust can temporarily be accumulated.} {We find that the 1 mm-sized dust remains in the co-orbital region of the planet as long as $1\gg St$. If the gas density in the co-orbital regions decreases enough for the Stokes number to reach about order unity, dust is removed from the co-orbital region and the entire gap empties out. The onset of this process can also be seen in \autoref{fig:rainbowplot} where dust in the co-orbital region is stable until about 120 planetary orbits. At that point, the depth of the gap edges reaches a temporary maximum (largest depth). After that, dust from the co-orbital region is lost and fills the gaps edges. The gap depth only grows again after all the dust from the co-orbital region is lost. This effect can be seen in all our simulations. However, it happens sooner in simulations containing a more massive planet or planets orbiting at smaller semi-major axes where the dynamical timescales are smaller.} {We also expect the onset of this process to occur sooner, the flux of dust through the gap to be increased and the peaks in the gap profile to be smoothed if dust turbulent diffusion is included \citep[see e.g. ][for a comparison with and without diffusion in 2D]{Zhu2012,Weber2018}.} The outer edge of the planet also acts as a barrier for 1 mm-sized dust particles drifting inwards \citep{Rice2006}. Hence, without diffusion, there is no inward drift from the outer part of the disk to the co-orbital region. While, for the Saturn-mass simulations, the dust is evenly distributed along the co-orbital radius, there are dust enhancements just before and after the location of the planet (around the Lagrange points L4 and L5) in the 1 Jupiter-mass simulations. This is also consistent with earlier work in 2D, e.g. \cite{Zhang2018}. \newline
For $M_p = 5M_{\text{Jup}}$, the dust surface density at the location of the planet is much lower than in the other simulations because dust is the co-orbital region is lost. The gap region in these simulations still contains dust, but it is not confined to the co-orbital radius. Instead, it is more evenly distributed across the gap, but at very low densities. Looking at the temporal evolution of these simulations, we find that the 5 $M_{\text{Jup}}$ simulations also produce distinct co-orbital dust accumulation features early in the simulations. However, dust originally trapped in this region vanishes after about 150 planetary orbits and is therefore not seen in \autoref{fig:surfacedensity_overview_dust} and \autoref{fig:surfdensprofiles_overview} where we plot a snapshot of the density distributions at 200 planetary orbits. We want to highlight again that the dust disk has a more complex temporal evolution than the gas disk without any sort of steady-state. Hence, the dust disk morphologies heavily depend on the age of the simulated systems. We expect the simulations containing a lower mass planet to also lose its dust from the co-orbital region at later times ($>$ 200 orbits). Moreover, for $M_p$ = 5 $M_\text{jup}$, we find the mass of the planet to be large enough to make the disk slightly eccentric, leading to non-axisymmetric gap structures \citep[][]{Kley2006,Szulagyi2017}.\newline
At $M_p$ = 5 $M_\text{jup}$ the surface density profile in gas (see \autoref{fig:surfdensprofiles_overview}) also has a W-shape. However, its origin is different from the situation explained above. Even though the dips adjacent to the planetary orbital radius become significantly deeper with the increase of the planetary mass from 1 $M_\text{jup}$ to 5 $M_\text{jup}$, the surface density at the orbital radius only decreases slightly. This is because gas efficiently accumulates in the {planet’s circumplanetary disk}. With the azimuthal averaging, the gas in the potential well contributes to the surface density profile at the planetary orbital radius even though, away from the planet, the gap is much deeper. \newline
When comparing the gap at different orbital distances, we find that the depth of the gap at 200 planetary orbits increases with planetary orbital radius. {One factor which favors gas gap formation at large radii in our models, is the adoption of constant kinematic viscosity. This becomes clear by studying the dimensionless P-parameter \begin{equation}
\label{eq:p-parameter}
    P = \frac{3}{4}\frac{h_g}{R_H}+\frac{50}{q\mathcal{R}}
\end{equation} 
introduced by \cite{Crida2005} which measures the ability of a planet to carve a gap in gas. The smaller this parameter, the easier it is to carve a gap. Here, $q=M_p/M_*$ is the planet-to-star mass ratio, $R_H=r_p(q/3)^{1/3}$ is the Hill radius of the planet and $\mathcal{R}=r^2\Omega/\nu$ is the Reynolds number. The first term on the r.h.s. of equation (\ref{eq:p-parameter}) scales as the aspect ratio $h_g/r$ and, for flared disks, increases with radius. With constant kinematic viscosity, the second term on the r.h.s. of equation (\ref{eq:p-parameter}) decreases as $r^{-1/2}$ i.e., it contributes less at large radii. In our models, the decrease of the second term, dominates over the increase of the first term. Therefore, the P-parameter decreases with radius and gap formation in gas is easier in the outer disk.}\newline
The outermost regions of the disk are depleted in dust. This is most noticeable for simulations with planets at large radii and small planet mass (e.g. \texttt{m50au1nep} in \autoref{fig:surfacedensity_overview_dust}). The dust depletion is due to radial inward drift of the 1 mm-sized dust and no replenishment through the closed outer radial boundary. We expect the entire outer disk to empty out eventually and drift inward to be trapped in a pressure maximum or at the inner boundary of our simulation domain.

\subsubsection{Vertical distribution}
We also investigate the vertical distribution of the gas and dust components. In \autoref{fig:verticaldistr_overview_gas} we plot the azimuthally averaged distributions of the volume density in gas. In \autoref{fig:verticaldistr_overview_dust} we plot the azimuthally averaged distributions of the volume density in dust. Both figures are plotted in a normalized logarithmic color-scale. The layout of the sub-panels in both figures is identical to \autoref{fig:surfacedensity_overview_gas} and \autoref{fig:surfacedensity_overview_dust}, i.e. the planet mass decreases from top to bottom and the orbital distance increases from left to right.\newline
Similarly to \autoref{fig:surfacedensity_overview_gas}, the gap in the vertical gas distribution is noticeable in all the simulations except in the Neptune-mass simulations. If a gap is present, it extends the full height of the disk. For more massive planets, the gap is wider at larger altitudes. For 1 $M_{\text{Jup}}$ planets and especially at 5 $M_{\text{Jup}}$ planets, an accumulation of gas at the center of the planetary potential is visible where gas {accumulates in the circumplanetary disk}. Since we underestimate the gravitational potential at the location of the planet due to the smoothing length, we expect this effect to be even more prominent in reality.\newline
The azimuthally averaged vertical density distribution in dust (\autoref{fig:verticaldistr_overview_dust}) is very different from the distribution in gas. This is mainly because dust is not pressure supported. The 1 mm-sized particles which are not perfectly coupled to the gas settle vertically. In a disk without a planet, all the 1 mm-sized dust does settle onto the midplane because we do not include turbulent diffusion in the simulations which would counteract the settling at some point. In \autoref{fig:verticaldistr_overview_dust}, the effect of settling is seen most clearly in the outermost regions of the disks containing a low-mass planet. There, the dust disk is very thin. Its thickness is not vanishingly small because the computational cells at the midplane have a non-vanishing size. \newline
{\autoref{fig:surfacedensity_overview_dust} shows that vertical settling is counterbalanced by vertical stirring around the gap edges, as predicted by \cite{Edgar2008}. The vertical stirring is caused by the meridional flows in the gas \citep[][]{Szulagyi2014,Fung2016,Szulagyi2021}. In the midplane where the mm-sized grains are most tightly coupled to the gas, they are dragged along with the vertical upward flow of the gas until they reach a height at which the grains decouple enough for them to settle toward the midplane again. The vertical stirring is not uniform along the entire edge (in azimuthal direction) but it is strongest at the location where the planetary wake meets the edge of the gap and decreases in strength further away from the planet. Due to the differential rotation of the disk, material at the gap edge has a different angular velocity than the planet. Therefore, in the rest frame of the gap edge, the vertical stirring is periodic with a period $t_\mathrm{stir}\simeq2\pi/\abs{\Omega_K(r_P)-\Omega_K(r_\mathrm{edge})}$ where $r_\mathrm{edge}$ is the radius of the gap edge.} Low-mass planets, like the Neptune-mass planets, only disturb the dust vertically in regions close to the orbit of the planet. The larger the planet mass, the thicker the dust disk becomes and regions farther away from the planetary orbit are affected. Similar effects were found by \cite{Fouchet2010} in their 3D gas+dust simulations but using the smoothed particle hydrodynamics (SPH) approach. In the simulations containing a larger mass planet, dust is present in a large vertical fraction of the simulation domain, also far away from the planet and the midplane. 
{We discuss the effects of turbulent diffusion on the vertical dust distribution in section \ref{sec:caveats}}.\newline
In some panels in \autoref{fig:verticaldistr_overview_dust}, dust gets puffed up at the inner computational boundary. This can be seen for example in the simulations containing a Neptune mass planet at 30 AU and 50 AU {and is caused by direct stellar irradiation of the inner edge of the disk.}

\subsection{Synthetic ALMA observations}
In this section, we study the synthetic ALMA mm-continuum observations created with {\scriptsize RADMC-3D} and {\scriptsize CASA}. \newline
\begin{figure*}
\centering
	\includegraphics[width=.85\textwidth]{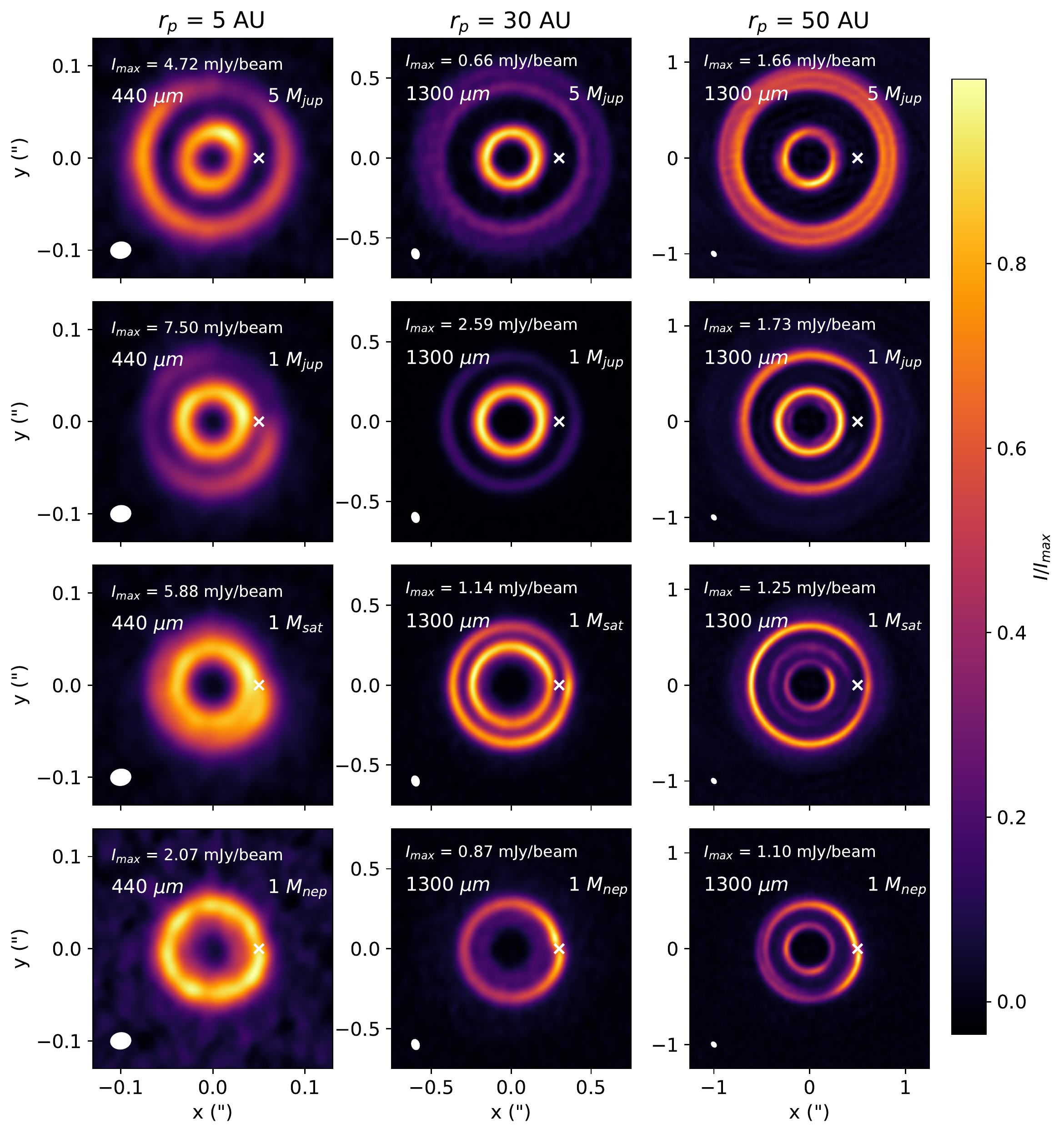}
\caption{For each of our 12 hydrodynamic simulations we show one synthetic ALMA observation. Each image is normalized to its peak intensity which is indicated in the upper-left corner of each image. From left to right we show the disk containing a planet at  $r_p$ = 5 AU, 30 AU, 50 AU. The rows show the disks with a planet of equal mass, $M_p$ = 5 $M_{\text{jup}}$, 1 $M_{\text{jup}}$, 1 $M_{\text{sat}}$, 1 $M_{\text{nep}}$. In each image, the mass of the planet is indicated in the upper right corner. The central wavelength of the observation band is indicated in the upper-left corner. The beam size  is indicated in the lower-left corner.}
\label{fig:synt_observation_overview}
\end{figure*}
In \autoref{fig:synt_observation_overview}, we show normalized mm-continuum maps for each of our 12 hydrodynamic simulations in the same layout as in \autoref{fig:surfacedensity_overview_gas}. For each simulation, we chose the observation band and antennae configuration such that it produces the most detailed intensity map. In detail, this is antenna configuration C43-8 and ALMA band 9 ($\sim$ 440 \micron) for simulations with $r_p$ = 5 AU and ALMA band 6 ($\sim$ 1300 \micron) for simulations with $r_p\geq$ 30 AU. The intensity maps are normalized with the peak intensity. We place the value of the peak intensity in the upper left corner of every individual map. The beam size  is indicated with a white ellipse in the bottom left corner of each map and the location of the planet is indicated with a white \texttt{x}. \newline
The intensity maps containing $M_p = 5M_{\text{Jup}}$ planets (first row in \autoref{fig:synt_observation_overview}) all show a clear gap with a wide outer ring and a narrower inner ring. At $r_p$ = 5 AU the inner and outer rings show asymmetric brightness distributions. The outer ring is brightest on the opposite side of the planet whereas the inner ring is brightest about 45 degrees in front of the planet. This is the location where the inner spiral arm meets the inner disk and stirs up the dust. {The asymmetry is also enhanced due to beam dilution, especially at $r_p$ = 5 AU, where the beam is quite large and elongated.} At $r_p\geq$ 30 AU, the outer rings are more azimuthally symmetric. Generally, the inner ring is brighter than the outer ring. \newline
{The outer ring in the simulation containing a $M_p = 5M_{\text{Jup}}$ planet at $r_p$ = 50 AU, shows two rings separated by a depletion. The origin of this feature can best be seen in the vertical density distribution in figure \ref{fig:aziavg_intensity} where the outer disk shows a vertical notch at around 1.7 $r_p$. This notch can also be seen in the vertical distribution of the disks containing a $M_p = 5M_{\text{Jup}}$ planet at $r_p$ = 30 AU and in the disk containing a $M_p = 1M_{\text{Jup}}$ planet at $r_p$ = 50 AU. However, in both other cases, the notch is not prominent enough to appear in the synthetic intensity map.}
One must be careful when interpreting the inner ring. We found that in all the $M_p = 5M_{\text{Jup}}$ maps, the inner ring is caused by an accumulation of dust at the inner computational boundary rather than at a gas pressure maximum. No matter where the inner boundary would lie, there would be naturally a ring in the inner boundary, where dust accumulates, due to the mass-conservation simulations (i.e. we do not allow outflow). In the maps containing lower-mass planets ($M_p < 5M_{\text{Jup}}$), {there are maps with either one or two inner rings. If only one ring is present, it has a physical origin. If there are two inner rings (e.g. Jupiter-mass, Saturn-mass and also Neptune mass at 50 AU), the outermost of the two inner rings is physical, the innermost ring is an artifact.} \newline
In the second row of \autoref{fig:synt_observation_overview} we show the $M_p = 1M_{\text{Jup}}$ planets which all produce two rings in the disk. {At $r_p$ = 5 AU, the inner ring has an asymmetric azimuthal brightness profile with a peak in front of the planet (in anticlockwise direction). At this location, the inner planetary wake meets the inner edge of the gap and stirs up dust above the midplane. Exposed to direct stellar radiation, dust above the midplane is warmer and, hence, brighter in the intensity map. The puffed up dust disk at this location also casts a shadow to the outer disk and the outer ring. Therefore, the outer ring is darker in the upper-right quadrant. Vertical stirring also occurs when the outer planetary wake meets the outer edge of the gap. Similarly to the inner ring, the outer ring is warmer and therefore brighter in the lower-right quadrant. As shown in \autoref{fig:verticaldistr_overview_gas}, stirring is much more effective at 5.2 AU compared to 30 AU and 50 AU because the mm-sized grains are more strongly coupled to the gas' meridional flows due to the larger gas density closer to the star. Therefore, at 30 AU and 50 AU less dust is stirred above the midplane which can cause shadowing. Hence, asymmetries due to absorption and shadowing are much less pronounced at larger orbital radii. Comparing the intensity map containing the $M_p = 1M_{\text{Jup}}$ planet at 30 AU and 50 AU it becomes apparent that the outer ring at 30 AU is dimmer than at 50 AU. This is because the Saturn-mass planet is able to stir up more dust in the inner disk than when it orbits at 50 AU. Again, this is due to better coupling to the gas. The puffed up inner edge of the gap, then casts a shadow over the vertically stirred dust at the outer ring. This shadow is slightly larger in the upper-right quadrant, but generally the brightness distribution is more symmetric than with the planet at 5 AU. When the Saturn-mass planet orbits at 50 AU, the inner edge of the gap casts a smaller shadow and the outer ring is brighter. Hence, even though the dust at the location of the outer ring is similar in both cases, the emission of the outer ring is different due to temperature differences as a result of the vertical structure of the disk.} {In all the $M_p = 1M_{\text{Jup}}$ maps,} the inner ring is clearly separated from our inner computational boundary.\newline
In the models containing the $M_p = 1M_{\text{Sat}}$ at $r_p$ = 5 AU it is difficult to resolve the gap due to the combination of small gap width and large beam size. {However, an asymmetry can be seen as in cases with more massive planets. Most prominent is the crescent-shaped asymmetry in the lower-right quadrant due to the vertical stirring at the location where the outer planetary wake meets the gap edge. The asymmetry arises because in the upper-right quadrant, the outer edge of the gap lies in the shadow of the puffed up inner edge section of the disk. Here, the asymmetries are also less pronounced at 30 AU and at 50 AU compared to when the planet orbits at 5 AU because grains are less strongly coupled to the meridional flows. Similarly to the case with $M_p = 1M_{\text{Jup}}$, the inner gap edge is less puffed up when the Saturn-mass planet orbits at 50 Au than at 30 AU. Therefore, the inner ring is less pronounced in the intensity map.} For $r_p\geq$ 30 AU, the gap is clearly resolved. At $r_p$ = 30 AU, there are two rings with roughly equal brightness. At $r_p$ = 50 AU, the ring to the inside of the gap is barely visible. Moreover, there is a third ring right at the inner boundary of our computational domain caused by accumulating dust. Also, faint emission from the disk outside the outer ring are visible. \newline
As mentioned in section \ref{sec:result_gas}, the Neptune mass planet barely opens a gap in the dust. There is no gap in the gas and hence, no pressure bump in which dust can accumulate. The features which we see in the disks containing a Neptune-mass planet are traces of the spiral wakes caused by the planet. The spiral wakes stir up dust from the midplane which gets illuminated and subsequently heated by the central star. 

\begin{figure*}
\centering
	\includegraphics[width=.7\textwidth]{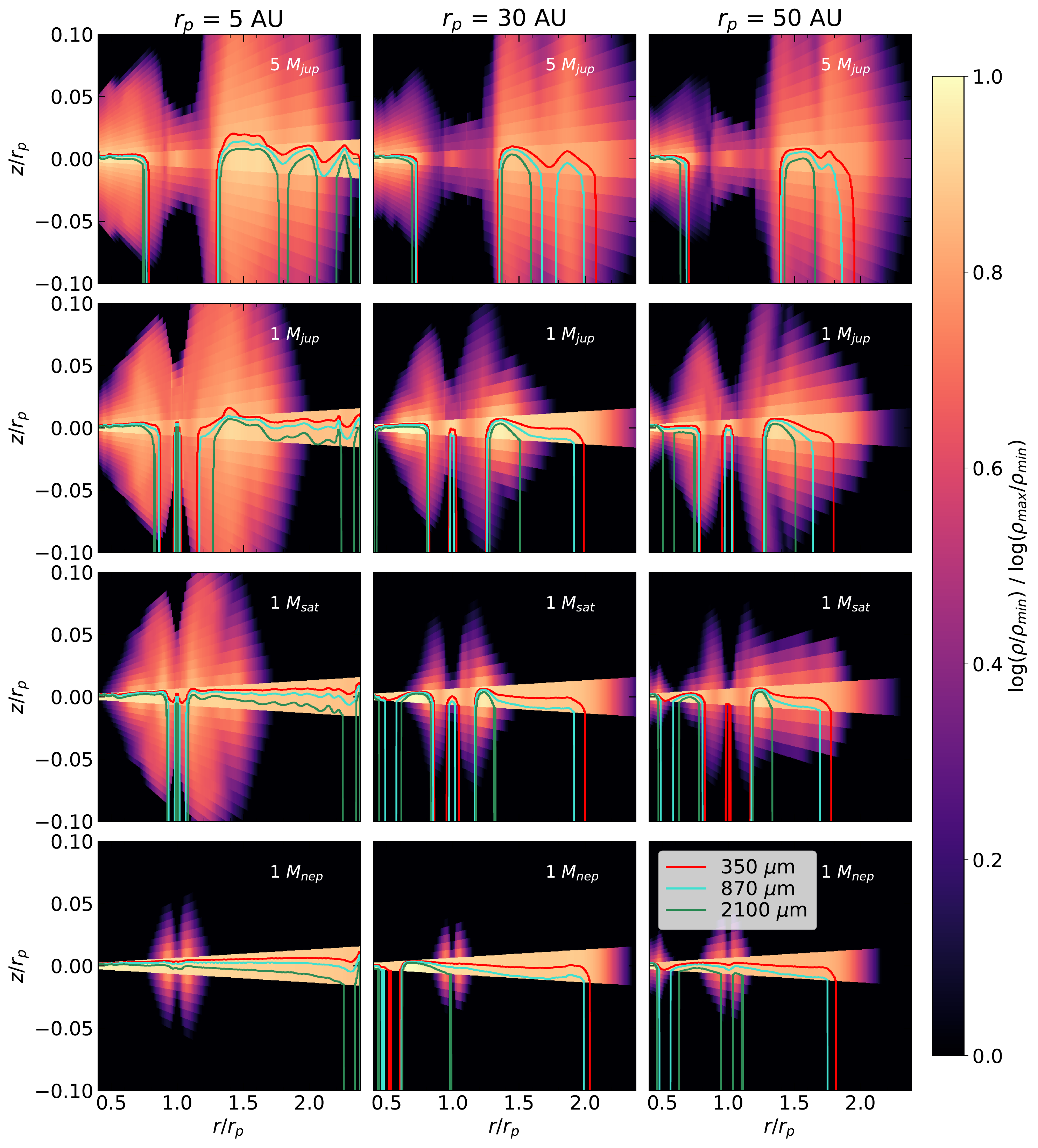}
\caption{This figure shows for each of our 12 models the contour $z_1(r)$ of the surface where the optical depth $\tau$ = 1. Each panel shows a vertical cut of the volume density in dust at the location of the planet ($\phi=0$). In each panel we show the $\tau$ = 1 contour for three wavelengths (350 \micron, 870 \micron, 2100 \micron). At each radius, we integrate along the z-axis from z = +$\infty$.}
\label{fig:tau1_surfaces}
\end{figure*}

\subsection{Gap-widths in gas}
Before we present our results on gap width measurements in the synthetic ALMA intensity maps, which we will do in the following section, we will present gap width measurements in gas. For this, we measured the gap widths $\Delta_g$ in the hydrodynamic gas density distribution. This allows us to better understand the results and lets us directly compare them to previous studies. Our approach for the gas gap measurements is identical to \cite{Zhang2018}. We find the following relation between the gas gap width $\Delta_g$ and model parameters: 
\begin{equation}
    \Delta_{g}=0.27\bigg(\frac{M_p}{M_*}\bigg)^{0.21}\bigg(\frac{h_g}{r}\bigg)^{-0.30} \alpha^{-0.15}
    \label{eq:K_Delta_gas}
\end{equation}
Compared to \cite{Kanagawa2016MassWidth}, the gap-width in gas $\Delta_{g}$ is, with a power law index 0.21, less sensitive to the planet-to-star mass-ratio ($0.5$ in \citealp{Kanagawa2016MassWidth}) but is in rough agreement with \cite{Zhang2018} who find a similar value ($0.26$). The gap width in gas is also less sensitive to $h_g/r$ and to $\alpha$ compared to \cite{Kanagawa2016MassWidth}. They find power law indices $-0.75$ and $-0.25$ respectively. The difference to \cite{Kanagawa2016MassWidth} is likely due to our different definition of the gap-width in gas. Our definition is identical to the definition in \cite{Zhang2018} which is smaller for wide gaps. Nevertheless, our gap width in gas is more sensitive to $h_g/r$ and $\alpha$ compared to what is found in \cite{Zhang2018}. {They find values of $-0.05$ and $-0.08$ respectively.} The differences likely arise due to a combination of 3-D and radiative effects.

\subsection{Gap widths in intensity maps}
We measure the widths of the gaps ($\Delta$) in azimuthally averaged intensity profiles (see \autoref{fig:aziavg_intensity}).
\subsubsection{Measuring gap widths}
\begin{figure}
	\includegraphics[width=0.9\columnwidth]{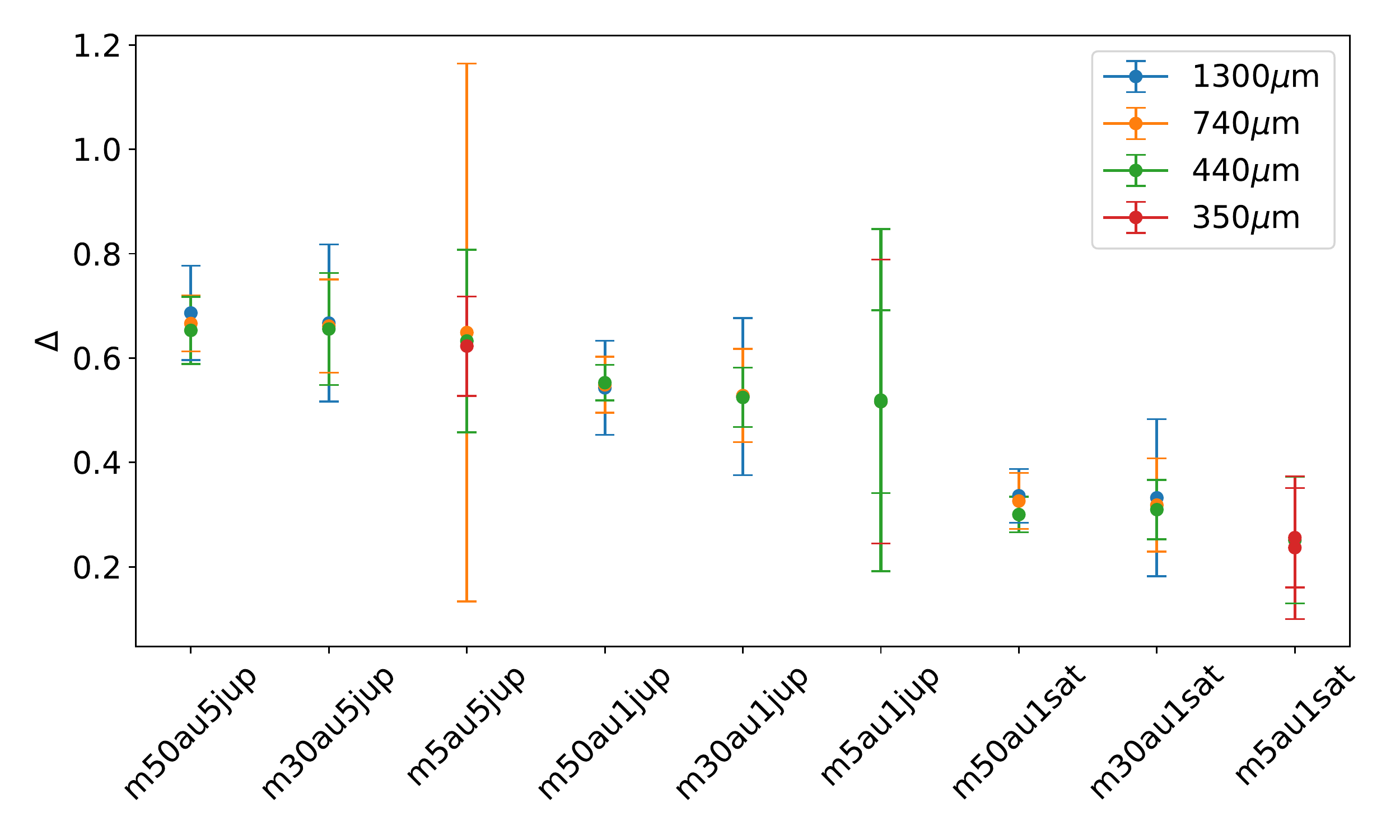}
\caption{Gap width measurements ($\Delta$) as defined in equation (\ref{eq:gap_width_definition}). In all our models which contains a planet massive enough to open an observable, we show the gap width measured in synthetic observations at different wavelengths. The vertical error bars correspond to the size of the beam in the synthetic ALMA observation.}
\label{fig:gap_delta_overview}
\end{figure}

We first identify the gap caused by a planet around $r_p$ and then find the first local maximum outside and inside the center of the gap. In the case of Neptune mass planets, which do not open observable gaps, we do not measure a gap width. In all other cases, we fit a Gaussian profile to the observed maxima and identify the radii at which the emission peaks as $r_{\mathrm{out}}$ and $r_{\mathrm{in}}$ respectively. We then define the gap width $\Delta$ as 
\begin{equation}
\label{eq:gap_width_definition}
    \Delta = \frac{r_{\mathrm{out}}-r_{\mathrm{in}}}{r_{\mathrm{out}}}.
\end{equation}
which has the advantage that it does not depend on $r_p$. In some cases, we can not identify a clear gap in our synthetic images but only a single ring located to the inside or to the outside of the planetary orbital radius. Then we assume the gap to be symmetric about the planetary orbital radius $r_p$, i.e., 
\begin{equation}
\label{eq:symmetric_assumption}
r_\mathrm{out}-r_p=r_p-r_\mathrm{in}.
\end{equation}
If the ring is located to the inside of the planetary radius, we use equation (\ref{eq:symmetric_assumption}) and express $r_{\mathrm{out}}=2r_p-r_{\mathrm{in}}$ which we then plug into the definition of the gap width (\ref{eq:gap_width_definition}) and use the value for $r_p$ which we know from the hydrodynamic simulations. If only an outer ring is visible, we use $r_{\mathrm{in}}=2r_p-r_{\mathrm{out}}$. Thus, we can always define a gap width $\Delta$ even if only a single ring is visible in the synthetic images. \newline
For the gap width measurements, we use the locations of the local maximum of the intensity to define a gap width instead of the locations of the edge at half of the peak value, as done by e.g., \cite{Zhang2018}. We found that the radial locations of the local maxima are less sensitive to the size of the beam. Furthermore, our definition is independent of the emission at the bottom of the gap which, for resolved gaps, can drop below the noise level. In \autoref{fig:gap_delta_overview}, we show the measured gap width $\Delta$ for all our models containing a planet more massive than Neptune. For each model, we measure the gap width ($\Delta$) at three different beam sizes. The vertical lines represent the size of the corresponding beam. We list the average values of the measured gap width in \autoref{table:fitting_results}. We later use these values in the fitting procedure in section \ref{sec:gapwidthfitting}.

\begin{figure}
	\includegraphics[width=.9\columnwidth]{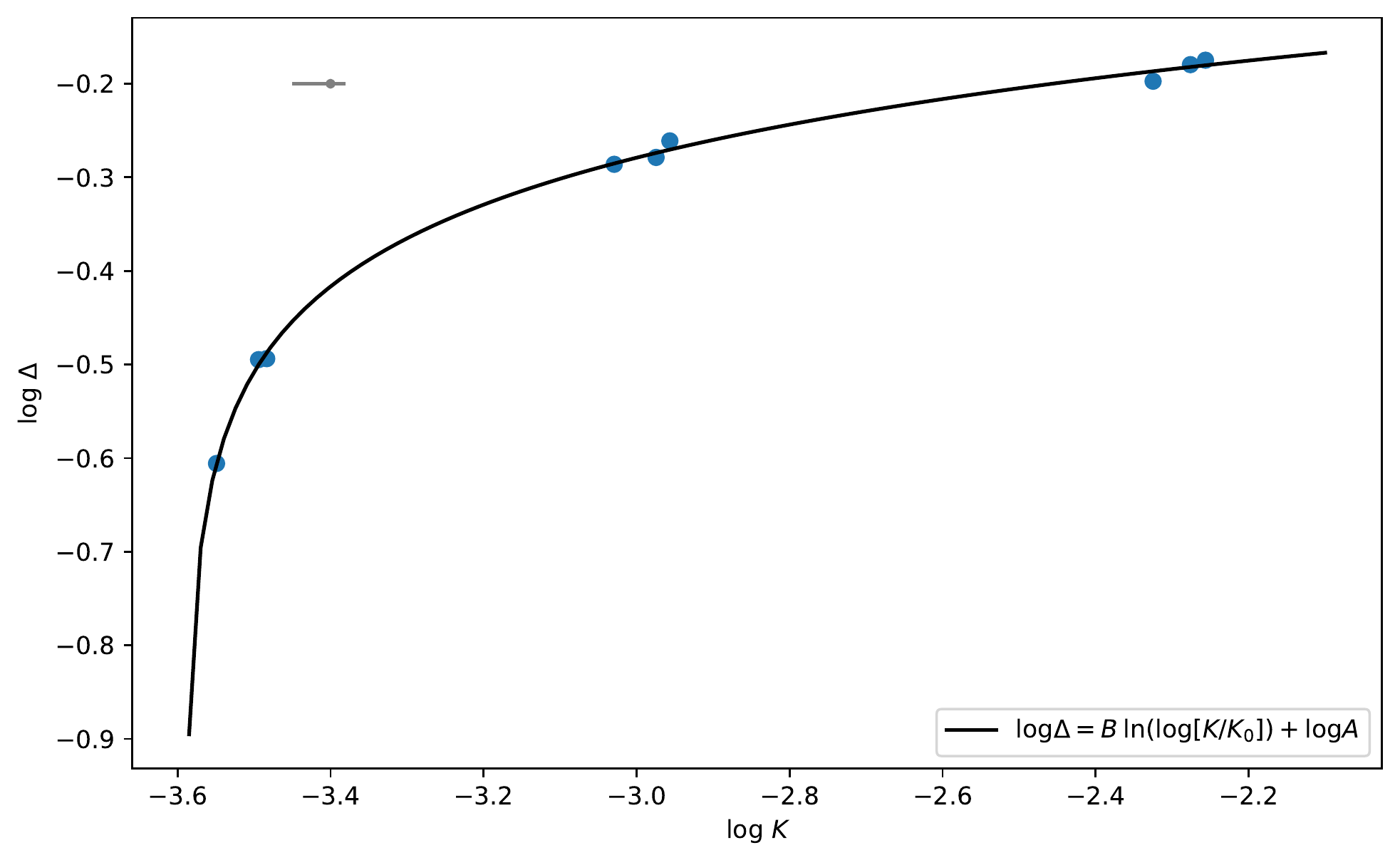}
\caption{$\Delta-K$-relation and best fit using equation (\ref{eq:delta_K_relation}). The gray error bar in the upper left corner shows the error in log K.}
\label{fig:gapwidthfit}
\end{figure}

\subsubsection{Fitting gap widths}
\label{sec:gapwidthfitting}
We derive an empirical relationship between the gap width $\Delta$, as measured from the intensity profiles in continuum images, and the planet mass $M_p$ similarly to what has been done in previous studies \citep[e.g.][]{Kanagawa2016MassWidth,Rosotti2016,Dong17,Zhang2018}. {While \cite{Kanagawa2016MassWidth} focused on the planetary gaps in gas, \cite{Dong17} focused on planet opened gaps in near-infrared scattered-light images. \cite{Zhang2018} have previously studied the relation between gap widths and planetary mass in continuum intensity maps. Our aim for this section is to do the analysis, for the first time, based on three-dimensional simulations, and to improve upon the previous approaches by providing a method tailored towards observations rather than hydrodynamic simulations.} Following previous studies, we define a dimensionless parameter $K$ which is proportional to the mass ratio between the planet and the star $M_p/M_*$, and has a power-law dependence on the parameters $H_B$ and $\alpha$ as
\begin{equation}
    K = \frac{M_p}{M_*}\cdot H_B^a\cdot\alpha^b.
    \label{eq:K_parameter}
\end{equation}
The dimensionless parameter $H_B$ can be regarded as an aspect ratio $H$ of the disk ($H= h_g/r$ and $h_g=c_s/\Omega $). We do not define $H_B$ in terms of hydrodynamic quantities in order to make this approach more tailored toward real observations. Therefore, we define $H_B$ as 
\begin{equation}
   H_B = \sqrt{\frac{\gamma k_B}{G M_* m_\mu}\cdot r T_B}.
   \label{eq:aspect_ratio_formula}
\end{equation}
Here, $k_B$ is the Boltzmann constant. The parameter $H_B$ is defined in a way that, for a vertically isothermal disk of temperature $T_B$ at radius $r$, $H_B=H$ holds. We determine the parameter $H_B$ for a given disk by measuring the (azimuthally averaged) brightness temperature $T_B$ of the dust emission at distance $r$ from the central star. We measure the brightness temperature $T_B$ instead of the physical temperature $T$ because, {similar to the aspect ratio of the disk},  we do not have direct observational access to the physical temperature in images of marginally optically thin regions. In optically thick regions, the brightness temperature is equal to the physical temperature of the emitting material at the $\tau=1$-surface. We choose $r=r_\mathrm{out}$ to be the radius of the ring outside to the planetary orbit $r_p$ where the dust emission is most optically thick and the brightness temperature approaches the physical temperature of the emitting material.\newline
The second dimensionless parameter in Eq. (\ref{eq:K_parameter}) is the Shakura and Sunyaev $\alpha$-parameter of turbulent viscosity \citep{Shakura1973}. {It is usally not directly measurable from mm-continuum observations but requires additional modeling as done by e.g. \cite{Pinte2016} for HL Tau.} Therefore, the value of the $\alpha$-parameter usually assumed to be in the range of $10^{-3}$ to $10^{-2}$ for typical disks. {The $\alpha$-turbulence parameter is also not a predetermined quantity in our radiative hydrodynamic simulations where we have adopted a constant kinematic viscosity instead of the alpha prescription typically used in isothermal hydrodynamic simulations.} Here, we estimate the $\alpha$-parameters from the scale height $h_g$ of a Gaussian fit to the vertical gas density in the initial hydrodynamic thermal equilibrium density field (without a planet) at the orbital radius of the planet $r_p$. {We do that before the injection of the planet because in our radiative disk the vertical gas profile can deviate from a Gaussian depending on the local heating and cooling in the gas. In addition to that, unlike in isothermal simulations, the local gas scale height sensitively depends on the radius at which it is measured due to the perturbation of the planet.} We then use the kinematic viscosity $\nu$ to compute the $\alpha$-parameter with the following formula: 
\begin{equation}
   \alpha = \frac{\nu}{\Omega \gamma h_g^2}
\end{equation}
We list the $\alpha$-parameters for each model in the third column of \autoref{table:fitting_results}.
\newline
In the next step, we find the best fitting parameters $a$ and $b$ in equation (\ref{eq:K_parameter}) which relate the $K$-parameter to the gap width $\Delta$ via the following relation
\begin{equation}
   \log \Delta = B\cdot \ln\big[\log \big(K/K_0\big)\big]+\log A
   \label{eq:delta_K_relation}
\end{equation}
where $A$, $B$ and $K_0$ are fitting coefficients. Here we highlight that we use the intensity profile of the synthetic continuum intensity maps to find the fitting parameters $a$ and $b$. This is in contrast to the previous work of \cite{Kanagawa2016MassWidth} and \cite{Zhang2018} who used the gas surface density to calibrate their parameters. We do not use the gas surface density in the fitting process here because we aim to produce a formula which is as independent as possible from hydrodynamic quantities. In equation (\ref{eq:delta_K_relation}), \textit{log} is the logarithm with base 10 and \textit{ln} is the natural logarithm. We chose an $\ln$-dependence between $\log \Delta$ and $\log K$ because we found it to be the functional dependence which minimizes the fitting error out of any monotonic functional dependence we tested (1st, 2nd order polynomial, exponential, root). {The $\ln$ functional dependence is also physically motivated because there exists a minimum planetary mass, which lies somewhere between a Neptune mass and a Saturn mass, below which no planetary gap is opened in a disk \citep[e.g.][]{Paardekooper2006}.} The $\ln$-dependence naturally provides a minimum value for $K$ below which a gap is not observable {unlike the linear dependence used in previous studies}. In equation \ref{eq:delta_K_relation} the lower bound is reached when $K = K_0$. \newline
In practice, we first find the best fitting coefficient $K_0$ for the case when $a=b=0$ in a least square fit. We find $K_0$ = $2.58\cdot10^{-4}$ which we fix at this value. It is equivalent to $\sim$5.2 $M_\mathrm{nep}$ which lies below a Saturn-mass. We do not yet fix the coefficients A and B in this process.  \newline
In a second step, {after fixing $K_0$}, we find the best fitting parameters $a$ and $b$ as defined in equation (\ref{eq:K_parameter}). For this, we assign values to $a$ and $b$ and perform additional least square fits with A and B as free parameters using the $\Delta-K$ relation in equation (\ref{eq:delta_K_relation}). We obtain a fitting error $\sigma$ from the sum of the square difference between the measured values of the gap width $\Delta$ and the fit. We then vary $a$ and $b$ to minimize $\sigma$. At the minimum $\sigma$, we find the optimal parameters $a$, $b$ and the corresponding coefficients A and B. Our results are A~=~0.61 and B~=~0.12 and the K-parameter is:
\begin{equation}
    K = \frac{M_p}{M_*}\cdot H_B^{0.086}\cdot\alpha^{-0.066}
    \label{eq:K_parameter2}
\end{equation}
We list the values of $\log K$ of each of the models in the fifth column of \autoref{table:fitting_results}. To compute the uncertainty of the fitting in $\log K$, we compute the residual between each measurement and the fitting curve. From the distribution, the lower bound is estimated by the 15.9 percentile, the upper bound by the 84.1 percentile. We find an uncertainty in $\log K$ of $^{+0.02}_{-0.05}$. We indicate this uncertainty with the gray error bar in the upper left corner of \autoref{fig:gapwidthfit}.\newline
{We can not immediately compare the gas relationship in equation (\ref{eq:K_Delta_gas}) with the observational relationship in equation (\ref{eq:delta_K_relation}) because of the different functional dependence (linear vs. $\ln$). However, because close to $\log K = -2.36$, the relation $\ln [\log (K/K_0)]\sim\log (K/K_0)$ holds, we can restrict ourselves to this value of $\log K$ and compare the power law exponents there. For example, the parameter $B$ provides the gap-width-dependence on the planet-to-star mass-ratio, the product $aB$ provides the dependence on $H_B$ or $h_g/r$ respectively. The product $bB$ provides the dependence on the $\alpha$-parameter. The dependence on $M_p/M_*$ is weaker in the observed gaps ($B=0.12$) than in gas ($B=0.21$). This is in agreement with \cite{Zhang2018} who also find a weaker dependence for moderately coupled grains. The dependence on $\alpha$ is also weaker in our observational relationship ($bB=-0.008$) compared to the dependence in gas ($bB=-0.15$). It is also about a factor two lower than for the moderately coupled large grains in \cite{Zhang2018} ($bB=0.016$). The dependence on $H_B$ is weaker and has the opposite sign ($aB=0.01$) than the dependence on $h_g/r$ in gas ($aB=-0.3$). Here, \cite{Zhang2018} find $aB=-0.01$ for their moderately coupled large grains. Hence, the magnitude of our value of $aB$ is in agreement with previous work, but it has opposite sign. This is a result of our definition of $H_B$ and the fact that the brightness temperature $T_B$ is not always a good tracer of the gas temperature. In our case, $T_B$ decreases faster with radius than the gas temperature, resulting in the opposite sign in the power law exponent of $H_B$ (to be precise, $rT_B$ is a decreasing function of radius, while in flared disks, $rT$ is generally an increasing function of radius for T being the physical gas temperature).} \newline
Equation (\ref{eq:delta_K_relation}) can be solved for the planet-to-star mass ratio $\log \big(M_p/M_*\big)$ from which we can directly compute planet masses $M_p$ from gap widths $\Delta$ in ALMA observations:
\begin{equation}
  \log \big(M_p/M_*\big) = \exp\big[\big(\log \Delta-\log A\big)/B\big]+\log\Big( K_0H_B^{-a}\alpha^{-b}\Big)
   \label{eq:planet_mass}
\end{equation}
Where A = 0.61, B = 0.12, $K_0$ = $2.58\cdot10^{-4}$, a = 0.086, b = -0.066.
We list the ratio $M_p/M_*$ found using equation (\ref{eq:planet_mass}) in the last column of \autoref{table:fitting_results}. We find a standard deviation of 10.1\% in the planet mass. Smaller deviations at smaller planet masses and larger deviations at larger planet masses. The maximum deviation is 21.9\%. {When comparing the mass ratios obtained with  equation (\ref{eq:planet_mass}), we generally find a good agreement with the mass ratio used in the hydrodynamic simulations. We find a better agreement for low-mass planets and worse agreement for high-mass planets when comparing to the actual value used in the simulation. This is likely due to the fact that the slope of the fitting function decreases with increasing planet mass (see figure \ref{fig:gapwidthfit}). Hence, small deviations in gap width measurements translate into larger mass deviation in the high-mass range compared to the low-mass range.} \newline

\begin{table*}
\begin{center}
\begin{tabular}{ l c c c c c c c} 
 \hline
  simulation &  $M_p/M_*$ & $\alpha$ & $T_B$ (K) &  $\log \Delta$ & $H_B$  & $\log K$  & $M_p/M_*$\\ 
  
            & (hydro)  & (hydro) & (measured) & (measured) & Eq. (\ref{eq:aspect_ratio_formula})     & Eq. (\ref{eq:K_parameter}) & Eq. (\ref{eq:planet_mass}) \\ 
  
  \hline
 \texttt{m5au1nep} & $5\cdot 10^{-5}$ & $2.1\cdot 10^{-2}$ & 14.1 & -- & 0.020 & -4.34 & --\\ 
 \texttt{m5au1sat} & $3\cdot 10^{-4}$ & $2.1\cdot 10^{-2}$ & 20.9 & -0.61 & 0.027 & -3.55 & $2.99\cdot 10^{-4}$\\ 
 \texttt{m5au1jup} & $1\cdot 10^{-3}$ & $2.1\cdot 10^{-2}$ & 15.2 & -0.29 & 0.024 & -3.03 & $0.99\cdot 10^{-3}$\\
 \texttt{m5au5jup} & $5\cdot 10^{-3}$ & $2.1\cdot 10^{-2}$ & 18.1 & -0.20 & 0.028 & -2.32 & $3.91\cdot 10^{-3}$\\ 
 \\
 \texttt{m30au1nep} & $5\cdot 10^{-5}$ & $4.4\cdot 10^{-3}$ & 5.6 & -- & 0.030 & -4.28 & --\\ 
 \texttt{m30au1sat} & $3\cdot 10^{-4}$ & $4.4\cdot 10^{-3}$ & 5.8 & -0.49 & 0.034 & -3.49 & $3.02\cdot 10^{-4}$\\ 
 \texttt{m30au1jup} & $1\cdot 10^{-3}$ & $4.4\cdot 10^{-3}$ & 4.4 & -0.28 & 0.031 & -2.97 & $0.94\cdot 10^{-3}$\\ 
 \texttt{m30au5jup} & $5\cdot 10^{-3}$ & $4.4\cdot 10^{-3}$ & 3.7 & -0.18 & 0.031 & -2.28 & $5.32\cdot 10^{-3}$\\ 
 \\
 \texttt{m50au1nep} & $5\cdot 10^{-5}$ & $4.5\cdot 10^{-3}$ & 5.4 & -- & 0.038 & -4.27 & --\\
 \texttt{m50au1sat} & $3\cdot 10^{-4}$ & $4.5\cdot 10^{-3}$ & 6.2 & -0.49 & 0.046 & -3.48 & $2.96\cdot 10^{-4}$\\
 \texttt{m50au1jup} & $1\cdot 10^{-3}$ & $4.5\cdot 10^{-3}$ & 6.6 & -0.26 & 0.052 & -2.96 & $1.12\cdot 10^{-3}$\\
 \texttt{m50au5jup} & $5\cdot 10^{-3}$ & $4.5\cdot 10^{-3}$ & 6.5 & -0.17 & 0.053 & -2.25 & $5.74\cdot 10^{-3}$\\ 
 \hline
\end{tabular}
\caption{We list the results of the gap width fitting. The second and third column list the planet mass and $\alpha$-turbulence parameter as found in the hydrodynamic simulations for each model. The fourth column lists the brightness temperature $T_B$ measured at the outer ring ($r_\mathrm{out}$) at a wavelength of $870\micron$ in the synthetic observations. The fifth column lists the gap widths $\log \Delta$ as measured in the azimuthally averaged intensity profiles. The sixth column lists, for each model, the parameter $H_B$ which we calculated using equation (\ref{eq:aspect_ratio_formula}). The seventh column lists the K parameter computed with equation (\ref{eq:K_parameter}) with the coefficients A, B, a and b as found in the fitting procedure. The last column lists the planet masses computed with equation (\ref{eq:planet_mass}) using the fitting coefficients found in the fitting procedure.}
\label{table:fitting_results}
\end{center}
\end{table*}

\subsection{Disk masses from synthetic observations}
Accurate accounting of bulk dust masses of protoplanetary disks is essential to understanding planet formation because it is the bulk reservoir of solid material from which terrestrial planets and the cores of giant planets form. There has been some discrepancy because recent observational studies \citep[e.g.][]{Andrews2013,Cieza2019b,Ansdell2016} reveal that the total dust mass ($\lesssim$cm sized grains) in Class II disks is low if one wants to explain the typical exoplanet demographics as derived from results of the \textit{Kepler} mission \citep[e.g.][]{Dong2013}. The shortage of solid material could be explained if the dust grains grow to larger sizes before the disk reaches the Class II stage. Larger grains remain hidden in the wavelength domains to which ALMA is sensitive. However, in this work, we focus on another explanation for the potential underestimation of the observed  mass of the solid disk components. \newline
Observationally, bulk dust masses are typically obtained {by} flux density measurements in (sub-)millimeter surveys and using the optically thin approximation (and making assumptions about opacity and temperature) \citep[e.g.][]{Tychoneiec2020}. Bulk dust masses can also be used as a proxy for bulk gas masses by assuming a dust-to-gas ratio. This indirect probe of bulk gas mass lends itself to the community because measuring bulk gas masses is difficult due to a lack of direct bulk gas mass tracers \citep[e.g.][]{BerginWilliams2017}. \newline
To probe dust masses, dust emissions are typically observed in (sub-)millimeter wavelengths. The observed intensity of optically thin emissions coming from an isothermal region of dust depends on the dust temperature $T_d$ and optical depth $\tau$. The optical depth is in turn dependent on the dust mass $M_d$ present along a line of sight and the opacity $\kappa$. As shown in \cite{Hildebrand1983}, the total dust mass $M_d$ in an isothermal region can be estimated from optically thin (sub-)millimeter emissions by measuring the flux density $F_\nu$ and using the following relation: 
\begin{equation}
M_d = \frac{F_\nu d^2}{\kappa_\nu B_\nu(T_d)}
\label{eq:dustmassequation}
\end{equation}
Here, $\kappa_\nu$ is the dust absorption opacity at the observed frequency $\nu$, $B_\nu$ is the Planck function at the dust temperature $T_d$ and $d$ is the distance between the emitting region and the observer. It is generally not easy to determine the opacity $\kappa$ and temperature $T_d$ of the emitting dust from observations. Hence, this approach of determining dust masses involves some uncertainties. However, we aim to apply this method here to investigate its validity and accuracy. As typically done in the community, we first approximate the dust opacity by a power law, 
\begin{equation}
\kappa_\nu=\: \kappa_0  \bigg(\frac{\nu}{\nu_0}\bigg)^\beta
\label{eq:dustopacity}
\end{equation}
\citep[e.g.][]{Draine2006} and use a simple parametrization as used by e.g. \cite{Beckwith1990AOBJECTS} or \cite{Ansdell2016} with $\kappa_{0}=10\: \mathrm{\frac{cm^2}{g}}$, $\nu_{0}=1000\: \mathrm{GHz}$ and power-law index $\beta=1$. Furthermore, we assume a characteristic disk temperature $T_{\text{dust}}$ = 20 K as in \cite{Ansdell2016} for all our models.
Using Eq. (\ref{eq:dustmassequation}), we compute the total dust mass $M_d$ from a synthetic image of each of our models. For this, we use synthetic images obtained at a wavelength of $\lambda$~=~870~$\micron$ (ALMA band 7). We list the results of this dust mass estimate in \autoref{table:dustmasses} column 4. At this wavelength the dust opacity, computed using Eq. (\ref{eq:dustopacity}), is $\kappa_\nu=3.45\:\mathrm{cm^2/g}$. For comparison, we list the actual dust mass present in the corresponding hydrodynamic model in the first third column of \autoref{table:dustmasses} and label it with $M_{d,\mathrm{hydro}}$. With this method, we recover, on average, 28.7\% of the total dust mass $M_{d,\mathrm{hydro}}$. The fraction is generally larger in models containing a planet at 5.2 AU and lower in models at 30 AU and 50 AU. One obvious reason for the deviation is the fact that the crude assumptions for temperature $T_d$ and opacity $\kappa_\nu$ are not perfectly representative of our models. The opacity used in the radiative transfer step at  $\lambda$~=~870 $\micron$ is with $\kappa_\nu=10.2\:\mathrm{cm^2/g}$ larger than what we assumed here. However, using a smaller opacity would lead to an overestimation of the recovered dust mass. \newline
We compute a characteristic disk temperature $\bar{T}_\text{dust}$ from the three-dimensional hydrodynamic temperature grid for each model, to assess the validity of the assumption of the characteristic temperature above. We compute $\bar{T}_\text{dust}$ as a mass-weighted average temperature
\begin{equation}
\bar{T}_\text{d} = \frac{1}{M_{d,\mathrm{hydro}}}\sum\limits_i m_{d,i}T_{d,i}
\label{eq:massaverage}
\end{equation}
where we sum over all the computational cells. Here, $m_{d,i}$ is the dust mass and $T_{d,i}$ is the dust temperature in cell $i$. As found by \cite{Ballering2019}, the radial extent of the disk has a large impact on the mass averaged dust temperature $\bar{T}_d$ and we find that smaller disks are significantly warmer than larger disks. The characteristic dust temperatures are listed for each disk model in \autoref{table:dustmasses} column 2. Except in \texttt{m5au1nep}, the characteristic dust temperatures are larger than 20 K in all the models containing a planet at 5.2 AU. In all other cases, the characteristic temperature is smaller as the disk extends farther away from the star. Characteristic temperatures below 20 K counteract the effect of the larger opacity used in our models compared to the initially assumed value of $\kappa_\nu=3.45\:\mathrm{cm^2/g}$. \newline
We repeat the dust mass calculations using the more suitable temperature and opacity values to compare with the crude estimates. We can only do this because we have the advantage over real observations to have perfect knowledge of opacity and temperature in our computational models. The results of this first improvement are listed in \autoref{table:dustmasses} column 5. On average, we recover only 30.3 \% of the total dust mass with this approach. The fact that we can not significantly increase the recovered dust mass points to a weakness in the applied approach itself and we point out several difficulties here. \newline
In the Rayleigh-Jeans tail of the Planck function, the sensitivity to changes in temperature increases the closer the peak wavelength is to the observed wavelength. At $T_d$ = 3.3 K, the peak of the Planck function is at the observed wavelength $\lambda$ = 870 $\micron$. Hence, characteristic temperatures close to 3.3 K introduce larger uncertainties in Eq. (\ref{eq:dustmassequation}) than larger temperatures. Hence, dust mass estimates will be more accurate for disks with larger characteristic temperatures and observations at longer wavelengths. Hence, dust mass estimates with $\bar{T}_\text{d}\sim3.3\:K$ rely on a very accurate estimation of the characteristic disk temperature. Generally, it is a difficult task to assign a single characteristic temperature to a disk because disks have diverse temperature profiles. Moreover, the assumption of optically thin emissions is not valid for large regions in the models. At a wavelength of 870 \micron, the emission becomes optically thick for surface densities $\Sigma_\text{d}>1/\kappa_\nu\sim0.1\:\mathrm{g/cm}^2$. In our initial conditions, this is the case in the inner disk for $r<$ 64 AU. Emissions at longer wavelengths are more optically thin \citep[e.g.][]{Liu2019,Zhu2019b}. At a longer wavelength of 1300~\micron, where we use $\kappa_\nu=7.8\:\mathrm{cm^2/g}$, the optically thick regions in the inner disk only {reach} up to $r<$~39~AU. As shown in \autoref{fig:surfdensprofiles_overview}, the surface density can also increase by a factor of a few in the rings at the edges of the disk gaps when the disks evolve away from their initial condition. Because optically thick emissions are independent of surface density and they lead to an underestimation of the bulk dust mass. Furthermore, the temperature in large regions of the models is so low that the emission at the observed wavelength is lost in the observational noise. Hence, dust in these regions remains undetected in the synthetic observations which further decreases the recovered dust mass fraction.\newline
We repeat the dust mass measurements at longer wavelengths of $\lambda$~=~1300~$\micron$ (ALMA band 6) and $\lambda$~=~2100~$\micron$ (ALMA band 4) (using identical estimates for $\bar{T}_\text{d}$ as before) and list the results in \autoref{table:dustmasses} column 6 and 7. Longer wavelengths are more favorable here because, firstly, the Planck function is less sensitive to temperature at longer wavelengths (in the Rayleigh-Jeans tail). Secondly, emission at longer wavelengths are more optically thin and thirdly, the cold outer regions of the disk emit at longer wavelengths. At $\lambda$~=~1300~$\micron$ we recover, on average, 30.2\% of the total dust mass and at $\lambda$~=~2100~$\micron$ we recover on average 33.7\% of the total dust mass. \newline
In this section, we applied a typical method used in the observational community to measure bulk dust masses by assuming optically thin emission and using equation (\ref{eq:dustopacity}). This allows us to evaluate the typical assumption of dust opacity and temperature made in previous studies. We showed a significant underestimation of the total dust mass, even under ideal conditions where opacity and dust temperature are well known. The assumption of optically thin emission fails in large fractions of the disk. Moreover, signal-to-noise ratios in the ALMA bands used here are not large enough to recover dust emissions from cold outer disk regions. Generally, the characteristic disk temperature we find increases with the mass of the planet. This is because a more massive planet stirs up more dust above the midplane which is then illuminated directly by the central star. This also leads to a trend that we recover more dust mass from disks with a low-mass planet. By assuming a constant $T_d$ = 20 K for all disks, we recover more dust from the disks containing a planet at a smaller semi-major axis than at a larger semi-major axis. By computing a characteristic dust temperature for every model individually, we see that the disks with a planet at larger radii are generally colder than 20 K and we do not see a significant difference in recovered disk mass for disks with planets at different radii anymore \citep[][]{Ballering2019}. Dust contained in the hidden optically thick regions in the disk can potentially account for the missing dust mass.\newline
Photon scattering can be an additional reason for the underestimation of the total dust mass. It decreases the maximum depth from which photons can escape and can make an optically thick region look optically thin \citep[e.g.][]{Rybicki1979}. The reduction of emission due to scattering is largely ignored in observations but can have an important impact \citep[e.g.][]{Zhu2019b}. \newline

\begin{table*}
\begin{center}
\begin{tabular}{l c c c  c c c c c c } 
 \hline
  simulation & $\bar{T}_\text{dust}$  & $M_{d,\text{hydro}}$ & $M_{d,\text{obs.}}$ &  $M_{d}$ & $M_{d}$ & $M_{d}$ & $M^{\tau<1}_{d}$ & $M_{d}^{\tau<1}$ & $M_{d}^{\tau<1}$ \\ 
  
  & (K)& ($M_{\earth}$) & ($M_{\earth}$) & ($M_{\earth}$) & ($M_{\earth}$) & ($M_{\earth}$) & ($M_{\earth}$) & ($M_{\earth}$) & ($M_{\earth}$) \\
  
   & &  & $\lambda=$870 \micron & $\lambda=$870 \micron &   $\lambda=$1300 \micron  & $\lambda=$2100 \micron & $\lambda=$870 \micron &   $\lambda=$1300 \micron  & $\lambda=$2100 \micron\\ 
  \hline
 \texttt{m5au1nep} & 10.8 & 4.95    & 1.33 & 1.27 & 1.44 & 1.72 & 1.82 & 2.35 & 3.21\\ 
 \texttt{m5au1sat} & 23.2 & 4.95    & 3.57 & 0.98 & 1.21 & 1.53 & 1.81 & 2.31 & 3.12\\ 
 \texttt{m5au1jup} &  25.7 & 4.95   & 3.66 & 0.87 & 1.07 & 1.35 & 1.67 & 2.16 & 2.86\\
 \texttt{m5au5jup} &  25.4 & 4.95   & 3.14 & 0.76 & 0.95 & 1.22 & 1.46 & 1.87 & 2.43\\ 
 \\
 \texttt{m30au1nep} & 3.7 & 68.5    & 3.39 & 78.0 & 50.8 & 39.0 & 43.1 & 50.3 & 54.9\\ 
 \texttt{m30au1sat} & 7.2 & 68.5    & 10.9 & 23.6 & 23.7 & 23.9 & 38.2 & 44.2 & 48.4\\ 
 \texttt{m30au1jup} & 8.5 & 68.5    & 12.7 & 20.3 & 21.9 & 23.2 & 34.8 & 41.2 & 47.6\\ 
 \texttt{m30au5jup} & 15.5 & 68.5   & 4.21 &  2.13 &  3.3 &  9.33 & 32.1 & 38.1 & 45.0\\ 
 \\
 \texttt{m50au1nep} & 5.1 & 147   & 8.17 & 57.7 & 51.5 & 53.1 & 94.9 & 113 & 128\\
 \texttt{m50au1sat} & 8.5 & 147   & 15.2 & 27.0 & 35.7 & 45.0 & 76.8 & 88.9 & 100\\
 \texttt{m50au1jup} & 9.5 & 147   & 29.2 & 38.5 & 50.8 & 65.6 & 72.3 & 85.0 & 102\\
 \texttt{m50au5jup} & 12.8 & 147  & 39.3 & 31.5 & 43.0 & 53.5 & 70.2 & 84.7 & 102\\ 
 \hline
\end{tabular}
\caption{This table gives total dust masses in the disks obtained from our synthetic observations using Eq. (\ref{eq:dustmassequation}). The second column lists mass weighted average temperatures of the three dimensional temperature field from {\scriptsize RADMC-3D}'s {\scriptsize mctherm}. The third column lists the total dust mass present in the hydrodynamic models. These are the reference values which we want to retrieve. The fourth column lists dust masses retrieved at 870 $\micron$ using Eq. (\ref{eq:dustmassequation}), identical dust temperature for all the models of $T_{\text{dust}}$ = 20 K and dust opacity $\kappa_\nu$ = 3.45 cm$^2$/g. Columns 5 to 7 contain total dust masses retrieved from using mass the weighted temperatures as listed in column 2 and opacities as used in the radiative transfer. In the models containing a planet at 5.2 AU, we used antennae configuration C43-10, for the other models we used antennae configuration C43-7. The last three columns contain the total dust mass above the $\tau=1$-surface, i.e. the optically thin dust mass, at different wavelengths, computed using the hydrodynamic dust density fields and the opacities as used in the radiative transfer.}
\label{table:dustmasses}
\end{center}
\end{table*}

\subsection{Optically thin dust and \texorpdfstring{$\tau=1$}{t=1}-surfaces}
In this section, we further explore the validity of the optically thin approximation and quantify the mass fraction of optically thin emitting dust. In \autoref{fig:tau1_surfaces}, we show a vertical cut of the dust volume density at the location of the planet ($\phi=0$) for each of our 12 hydrodynamical models. Overplotted are the contours $z_1(r)$ of the surfaces where the optical depth equals unity ($\tau$ = 1) when integrated along the z-axis from z = +$\infty$ for three different wavelengths (350 \micron, 870 \micron, 2100 \micron), i.e.:
\begin{equation}
    \tau=\kappa_\nu\int_\infty^{z_1}\rho_d dz
\end{equation}
All the areas below the contours are optical thick regions and hidden in face-on observations at the corresponding wavelengths.\newline
We computed the azimuthally averaged height $z_1$ of the $\tau=1$-surface at every radius and used it to compute the total dust mass which is optically thin$M_{d}^{\tau<1}$. The results at different wavelengths are listed in columns 8 to 10 of \autoref{table:dustmasses}. Assuming perfect knowledge of the emission $F_\nu$ and temperature $T_d$, we would be able to also obtain these values using equation (\ref{eq:dustmassequation}) as done in the previous section. Similar to the previous section, the total optically thin dust mass is larger at longer wavelengths and large radii because the dust is less optically thick at longer wavelengths and larger radii.\newline
Even at longer wavelengths, a large mass fraction of the disk remains optically thick in all the models. Dust in optically thick layers of the disk remains hidden and can not be recovered when doing bulk mass estimates using the optically thin assumption. The optically thin disk regions are mainly the gap regions and the outermost disk regions where the dust density has decreased due to radial inward drift. 

\section{Discussion}
\label{sec:discussion}
\subsection{Dust temperature}
The method with which we compute the gas temperature is different from the method with which we compute the dust temperature. We compute the gas temperature self consistently during the {radiative hydrodynamics} simulations based on local cooling and heating. The dust temperature, on the other hand, we compute with {\scriptsize RADMC-3D}'s \texttt{mctherm} in a post-processing step. The resulting {mass-weighted averaged dust} temperatures are {are 25\% larger in the disks containing a planet at 50 AU, 40\% lower in the disks containing a planet at 30 AU and 41\% lower in the disks containing a planet at 5.2 AU.} To investigate the validity of our approach, we ran supplementary computations with {\scriptsize RADMC-3D}'s \texttt{mctherm} and \texttt{image} tasks. In these test runs, we added a second dust species consisting of small (1 $\micron$ sized) dust particles. We assumed them to be well mixed with the gas with a local density $\rho_\mathrm{small}=10^{-4}\rho_g$, i.e. the global mass ratio between the small and the 1 mm-sized grains is 0.01. For the small grains, we computed an additional opacity table using Mie theory and the BHMIE code. In this two species setup, the upper disk layers are mainly filled with a small amount of $\micron$-sized particles while the midplane is dominated by the larger mm-sized grains as is expected in real disks. 
The resulting dust temperature {distribution} of the mm-sized grains computed with the \texttt{mctherm} task using this setup was similar to the one-species setup. {However, the midplane dust temperature was somewhat larger due to more indirect radiation coming from the upper disk layers where the small grains are.} Even though this approach probably represents a more realistic disk, it did not have any major effect on our results presented in this paper. 
\subsection{Caveats}
\label{sec:caveats}
There are some caveats regarding the assumptions used in our models. Firstly, we investigate the observable disk features after 200 planetary orbits. i.e. a snapshot in time, we do not follow a longer disk evolution. Dust-included simulations are continuously evolving, it is well known that they could never reach steady-state, due to the nature of the dust-gas interaction. Therefore, the disk sub-structures, and especially the gap widths, somewhat change with time as mentioned in section~\ref{sec:result_gas}. Hence, the time of observation has a crucial impact on the resulting observed gap width. However, we usually only know little about the time when planets formed in disks. \cite{Zhang2018} have done some analytical estimates on how the gap width changes with time. They estimate that for marginally coupled dust particles ($St\gtrsim10^{-2}$), the gap width is proportional to $St\times t$. This means that particles will drift twice farther over twice the amount of time. A detailed study which includes the change of the gap width over time is necessary to fully investigate this behavior.\newline  
Secondly, we only include one dust fluid in our simulations which represents 1 mm-sized dust. {Also, it should be stressed that the empirical fit, found in section \ref{sec:gapwidthfitting}, was found based on only one dust fluid.} This limitation is due to the enormous computational time needed for multi-fluid simulations of a global disk and perturbed by a planet in three dimensions. Of course, in reality, the dust consists of a distribution of grain-sizes. Dust-continuum observations at a given wavelength are most sensitive to emissions of a single particle size{. However,} larger and smaller sizes also contribute \citep[see e.g.][]{Draine2006}. The populations of larger/smaller particles are more weakly/strongly coupled to the gas which will result in a different spatial distribution in the disk. Hence, we expect slightly different intensity distributions \citep[e.g.][]{Drazkowska2019}. Including additional dust particle sizes is planned for future studies, when computers can handle such heavy computations in 3D. \newline
Thirdly, we do not include turbulent diffusion in the dust. {In vertical direction, dust turbulent diffusion is responsible for the mixing of dust particles and can counterbalance vertical settling toward the midplane \citep[][]{Dubrulle1995}. However, in this study, we focus on vertical mixing by a planet only. An estimate of the scale height of dust grains when turbulent diffusion is included is given by \citealp{Youdin2007}}: 
\begin{equation}
    h_d \approx h_g\sqrt{\frac{\alpha}{\alpha+St}\bigg(\frac{1+St}{1+2St} \bigg)}.
    \label{eq:dust_scale_height}
\end{equation}
{where the Stokes number $St$ is typically evaluated at the midplane. In our simulations, before inserting the planet, the dust scale height calculated with equation (\ref{eq:dust_scale_height}) is $\sim0.29\:h_g$ at 50 AU which is $\sim$2.0 times the vertical height of one grid cell. At 30 AU $\sim0.39\:h_g$ ($\sim$2.4 grid cells) and at 5.2 AU it is $\sim0.80\:h_g$ ($\sim$3.8 grid cells). The vertical extent of intermediately coupled grains ($St<1$) is typically not Gaussian, but follows a flatter distribution with a sharp cut-off due to the grains decoupling in the low gas density regions above the midplane \citep[e.g.][]{Fromang2009}. {This is because grains tend to decouple in the low gas density regions and vertical downward settling becomes dominant over vertical upward diffusion.} Hence, we expect only very few dust grains above $h_d$ {without additional stirring by a planet. Therefore, we expect dust which is stirred up significantly above $h_d$ by a planet, to be only marginally affected by vertical diffusion because its dynamics is dominated by planetary stirring and vertical settling.}\newline
The finite thickness of the midplane in the vertical dust distribution in our simulations due to the finite size of the vertical grid spacing, is a crude representation of a flat distribution with a sharp cutoff. However, {as stated above,} it is a factor 2 to 4 thinner than when including dust turbulent diffusion. \newline
{We measured the ratio $h_d/h_g$ in our simulations by fitting a Gaussian profile to the vertical density profiles and found that Jupiter-mass planets and 5 Jupiter-mass planets, at all orbital distances, are able to locally increase $h_d/h_g$ to values larger than what is expected from equation (\ref{eq:dust_scale_height}) i.e., by turbulent diffusion only. We expect this to also be possible for Saturn-mass planets (and possibly below) if the magnitude of turbulent diffusion is decreased. In the case of the 5 Jupiter-mass planets, the peak value $h_d/h_g$ is on average $0.75$. In the case of the Jupiter-mass planets, it is $0.84$ and in the case of Saturn-mass planets it is $0.23$. In the case of Neptune-mass planets, the vertical scale height is smaller than what we can resolve.}\newline
Studying the effects of dust turbulent diffusion will be a follow-up study to this project. {Besides its influence on the vertical distribution,} we expect dust turbulent diffusion to also affect the dust surface density distribution in a way that it smears out small-scale features, with a possible impact on gap width measurements. We ran an additional simulation identical to \texttt{m30au1jup} but with dust turbulent diffusion included as described in appendix A of \cite{Weber20}. There, we found a difference in gap width $\Delta$ between the cases with and without dust turbulent diffusion of 7\%.} \newline
Furthermore, in our setup, there is no thermal coupling between dust and gas. The temperature of the gas as determined in the {radiative hydrodynamics} simulations is independent of the dust temperature determined with the Monte Carlo approach. Thermal coupling is generally strongest in regions where the gas density is large, and the dust temperature is low, which is most likely the case in the midplane of the disk \citep[e.g.][]{Armitage2009,Voroboyov2020}. \newline
Moreover, in our hydrodynamical simulations, there were no magnetic fields included. This could also heavily alter the dust and gas distributions.
\newline
{We do not consider planetary migration in our simulations, even though gap opening planets can undergo type II migration. However, the migration timescales of type II migration is on the order of the viscous timescale \cite{Durmann15}. As pointed out in section \ref{sec:result_gas}, the viscous timescale across the length scale of a gap is considerably larger than the duration of our simulations. Hence, we do not expect type II migration to play a role. The Neptune-mass planets in our models can undergo type I migration because they perturb the gas disk only little and do not open a gap in gas. However, the type I migration timescale of a Neptune-mass planet is on the order of 100 times longer than the simulation time considered here \cite{Tanaka2002}. Therefore, we also expect type I migration to not change our results. A potentially relevant effect is the rapid type III runaway migration \citep{Masset2003}. This type of planetary migration is especially relevant for the Saturn-mass planets in our models, which are in the transition region between type I and type II migration. \cite{Masset2003} found that, for a disk with aspect ratio 0.05, which is approximately the case for our disk at 50 AU, and kinematic viscosity identical to our value at 50 AU, a Saturn-mass planet undergoes type III migration if the Toomre Q is below a value of $\sim10$. At 50 AU, we find $Q\sim5$. Hence, the Saturn-mass planet at 50 AU can potentially undergo type III migration. At 30 AU, the Toomre Q is larger, but at the same time, the aspect ratio is smaller. We expect type III migration to be less likely for the Saturn-mass planet at 30 AU. The Jupiter-mass planet at 50 AU is close to the transition region between type III and type II migration. If a planet indeed undergoes type III migration, \cite{Masset2003} find different migration behavior depending on the slope of the gas surface density. For shallow profiles, as in our case, they find a 50\% increase of the planet's semi-major axis, i.e., outward migration, within 50 orbital periods. {This is much shorter than the duration of our simulations.}. Hence, we expect rapid type III migration to affect the gap structure in the disk and should be taken into account in future studies.}

\section{Conclusions}
\label{sec:conclusions}
Using three-dimensional two-fluid hydrodynamic simulations of circumstellar disks with an embedded planet, we investigate observable planetary features in synthetic (sub-)mm-continuum ALMA images of the disks. We chose the grain size to be 1 mm within the hydrodynamic simulations. The feedback of the dust onto the gas is included in our simulations. We specifically investigate the gap widths caused by planets of different mass (Neptune-, Saturn-, Jupiter, 5 Jupiter-mass) at different orbital distances to the central star (5.2 AU, 30 AU, 50 AU). We summarize our results in the following points. 
\begin{itemize}
    \item Except for the Neptune-mass planets, the planets in our disk model open an annular gap at their orbital radius in both the dust and the mm-sized dust. The Neptune mass-planet can not disturb the gas enough to open a gap in the gas. In the dust, the Neptune-mass planets barely open a gap. 
    
    \item The temporal evolution of the surface density profile in dust is distinctly different from the surface density profile in gas. Whereas gap widths in both gas and dust steadily increase with time, the depth of the gap steadily increases with time only in the gas. The depth of the gap in dust can also decrease again with time after an initial increase in depth. 
    
    \item The planets cause significant vertical stirring of the dust which opposes the vertical settling. This creates thicker dust disks than in disks without a planet. The amount of vertical stirring depends on the mass and orbital radius of the planet. Large dust particles in the upper layers of the disk potentially have observational consequences. We examine this effect further in a follow-up paper. 
    
    \item We find multiple rings in the synthetic ALMA images which are caused by dust concentrations at the edges of the planetary gaps. 

    \item We examined the relation between the gap width as observed in ALMA images and the planet mass. We fitted the results and created equations between the planetary mass and the ALMA gap width, based on the planetary orbital radius, the disk turbulence and disk temperature. This relation can be used to constrain the planetary mass in future ALMA observations of gaps in circumstellar disks.
    
    \item We derived the disk mass from the hydrodynamical simulations and from the ALMA mock images created from the same simulations (\autoref{table:dustmasses}). We found a significant difference between the disk masses in the two cases: using the usual disk mass formula with optically thin dust emission assumption greatly underestimates the disk mass. The discrepancy usually a factor of few, highlighting the circumstellar disks might be several (up to 10) times more massive than previously thought. This has a strong consequence on planet formation, and disk processes, including chemistry.  Further, we found that the derived disk masses were generally larger in disks containing a low-mass planet, regardless of the orbital distance between the star and the planet. 
\end{itemize}

\section*{Acknowledgements}

J.Sz. thanks for the financial support through the Swiss National Science Foundation (SNSF) Ambizione grant PZ00P2\_174115. These results are part of a project that has received funding from the European Research Council (ERC) under the European Union’s Horizon 2020 research and innovation programme (Grant agreement No. 948467). Computations partially have been done on the "Piz Daint" machine hosted at the Swiss National Computational Centre. T.B. acknowledges funding from the European Research Council (ERC) under the European Union's Horizon 2020 research and innovation programme under grant agreement No 714769. F.B. and T.B acknowledge funding from the Deutsche Forschungsgemeinschaft under Ref. no. FOR 2634/1 and under Germany's Excellence Strategy (EXC-2094–390783311).

\section*{Data availability}
The data underlying this article will be shared on reasonable request to the corresponding author.




\bibliographystyle{mnras}

\begin{thebibliography}{}
\makeatletter
\relax
\def\mn@urlcharsother{\let\do\@makeother \do\$\do\&\do\#\do\^\do\_\do\%\do\~}
\def\mn@doi{\begingroup\mn@urlcharsother \@ifnextchar [ {\mn@doi@}
  {\mn@doi@[]}}
\def\mn@doi@[#1]#2{\def\@tempa{#1}\ifx\@tempa\@empty \href
  {http://dx.doi.org/#2} {doi:#2}\else \href {http://dx.doi.org/#2} {#1}\fi
  \endgroup}
\def\mn@eprint#1#2{\mn@eprint@#1:#2::\@nil}
\def\mn@eprint@arXiv#1{\href {http://arxiv.org/abs/#1} {{\tt arXiv:#1}}}
\def\mn@eprint@dblp#1{\href {http://dblp.uni-trier.de/rec/bibtex/#1.xml}
  {dblp:#1}}
\def\mn@eprint@#1:#2:#3:#4\@nil{\def\@tempa {#1}\def\@tempb {#2}\def\@tempc
  {#3}\ifx \@tempc \@empty \let \@tempc \@tempb \let \@tempb \@tempa \fi \ifx
  \@tempb \@empty \def\@tempb {arXiv}\fi \@ifundefined
  {mn@eprint@\@tempb}{\@tempb:\@tempc}{\expandafter \expandafter \csname
  mn@eprint@\@tempb\endcsname \expandafter{\@tempc}}}

\bibitem[\protect\citeauthoryear{{ALMA Partnership} et~al.,}{{ALMA Partnership}
  et~al.}{2015}]{ALMApartnership2015}
{ALMA Partnership} et~al., 2015, \mn@doi [Astrophysical Journal Letters]
  {10.1088/2041-8205/808/1/L3}, 808, L3

\bibitem[\protect\citeauthoryear{Andrews, Rosenfeld, Kraus  \& Wilner}{Andrews
  et~al.}{2013}]{Andrews2013}
Andrews S.~M.,  Rosenfeld K.~A.,  Kraus A.~L.,   Wilner D.~J.,  2013, \mn@doi
  [The Astrophysical Journal] {10.1088/0004-637X/771/2/129}, 771, 129

\bibitem[\protect\citeauthoryear{Andrews et~al.,}{Andrews
  et~al.}{2018}]{Andrews2018}
Andrews S.~M.,  et~al., 2018, \mn@doi [The Astrophysical Journal]
  {10.3847/2041-8213/aaf741}, 869, L41

\bibitem[\protect\citeauthoryear{Ansdell et~al.,}{Ansdell
  et~al.}{2016}]{Ansdell2016}
Ansdell M.,  et~al., 2016, \mn@doi [The Astrophysical Journal]
  {10.3847/0004-637x/828/1/46}, 828, 46

\bibitem[\protect\citeauthoryear{Armitage}{Armitage}{2009}]{Armitage2009}
Armitage P.~J.,  2009, {Astrophysics of planet formation}.
Cambridge University Press

\bibitem[\protect\citeauthoryear{Avenhaus et~al.,}{Avenhaus
  et~al.}{2018}]{Avenhaus2018}
Avenhaus H.,  et~al., 2018, \mn@doi [The Astrophysical Journal]
  {10.3847/1538-4357/aab846}, 863, 44

\bibitem[\protect\citeauthoryear{Bae, Zhu  \& Hartmann}{Bae
  et~al.}{2017}]{Bae2017}
Bae J.,  Zhu Z.,   Hartmann L.,  2017, \mn@doi [The Astrophysical Journal]
  {10.3847/1538-4357/aa9705}, 850, 201

\bibitem[\protect\citeauthoryear{Ballering \& Eisner}{Ballering \&
  Eisner}{2019}]{Ballering2019}
Ballering N.~P.,  Eisner J.~A.,  2019, \mn@doi [The Astronomical Journal]
  {10.3847/1538-3881/ab0a56}, 157, 144

\bibitem[\protect\citeauthoryear{Barge, Ricci, Carilli  \&
  Previn-Ratnasingam}{Barge et~al.}{2017}]{Barge2017}
Barge P.,  Ricci L.,  Carilli C.~L.,   Previn-Ratnasingam R.,  2017, \mn@doi
  [Astronomy and Astrophysics] {10.1051/0004-6361/201629918}, 605

\bibitem[\protect\citeauthoryear{Beckwith, Chini  \& G{\"{u}}sten}{Beckwith
  et~al.}{1990}]{Beckwith1990AOBJECTS}
Beckwith S. V.~W.,  Chini R.~S.,   G{\"{u}}sten R.,  1990, The Astronomical
  Journal, 99, 924

\bibitem[\protect\citeauthoryear{Ben{\'{i}}tez-Llambay, Krapp  \&
  Pessah}{Ben{\'{i}}tez-Llambay et~al.}{2019}]{Benitez-Llambay2018}
Ben{\'{i}}tez-Llambay P.,  Krapp L.,   Pessah M.~E.,  2019, The Astrophysical
  Journal Supplement Series, 241, 25

\bibitem[\protect\citeauthoryear{Bergin \& Williams}{Bergin \&
  Williams}{2017}]{BerginWilliams2017}
Bergin E.~A.,  Williams J.~P.,  2017, in , Formation, Evolution, and Dynamics
  of Young Solar Systems.
Springer, pp 1--38

\bibitem[\protect\citeauthoryear{Bohren \& Huffman}{Bohren \&
  Huffman}{1984}]{Bohren1984}
Bohren C.~F.,  Huffman D.~R.,  1984, Nature, 307, 575

\bibitem[\protect\citeauthoryear{Bouchut, Jin  \& Li}{Bouchut
  et~al.}{2003}]{Bouchut2003}
Bouchut F.,  Jin S.,   Li X.,  2003, \mn@doi [SIAM Journal on Numerical
  Analysis] {10.1137/S0036142901398040}, 41, 135

\bibitem[\protect\citeauthoryear{Cieza et~al.,}{Cieza
  et~al.}{2019}]{Cieza2019b}
Cieza L.~A.,  et~al., 2019, \mn@doi [Monthly Notices of the Royal Astronomical
  Society] {10.1093/mnras/sty2653}, 482, 698

\bibitem[\protect\citeauthoryear{Commer{\c{c}}on, Teyssier, Audit, Hennebelle
  \& Chabrier}{Commer{\c{c}}on et~al.}{2011}]{Commercon2011}
Commer{\c{c}}on B.,  Teyssier R.,  Audit E.,  Hennebelle P.,   Chabrier G.,
  2011, \mn@doi [Astronomy and Astrophysics] {10.1051/0004-6361/201015880},
  529, A35

\bibitem[\protect\citeauthoryear{Crida, Morbidelli  \& Masset}{Crida
  et~al.}{2006}]{Crida2005}
Crida A.,  Morbidelli A.,   Masset F.,  2006, \mn@doi [Icarus]
  {10.1016/j.icarus.2005.10.007}, 181, 587

\bibitem[\protect\citeauthoryear{Dipierro \& Laibe}{Dipierro \&
  Laibe}{2017}]{Dipierro2017}
Dipierro G.,  Laibe G.,  2017, \mn@doi [Monthly Notices of the Royal
  Astronomical Society] {10.1093/mnras/stx977}, 469, 1932

\bibitem[\protect\citeauthoryear{Dipierro, Laibe, Price  \& Lodato}{Dipierro
  et~al.}{2016}]{Dipierro2016}
Dipierro G.,  Laibe G.,  Price D.~J.,   Lodato G.,  2016, \mn@doi [Monthly
  Notices of the Royal Astronomical Society: Letters] {10.1093/mnrasl/slw032},
  459, L1

\bibitem[\protect\citeauthoryear{Dipierro et~al.,}{Dipierro
  et~al.}{2018}]{Dipierro2018}
Dipierro G.,  et~al., 2018, \mn@doi [Monthly Notices of the Royal Astronomical
  Society] {10.1093/mnras/sty181}, 475, 5296

\bibitem[\protect\citeauthoryear{Dong \& Zhu}{Dong \& Zhu}{2013}]{Dong2013}
Dong S.,  Zhu Z.,  2013, \mn@doi [Astrophysical Journal]
  {10.1088/0004-637X/778/1/53}, 778

\bibitem[\protect\citeauthoryear{Dong, Zhu, Rafikov  \& Stone}{Dong
  et~al.}{2015}]{Dong2015}
Dong R.,  Zhu Z.,  Rafikov R.~R.,   Stone J.~M.,  2015, \mn@doi [Astrophysical
  Journal Letters] {10.1088/2041-8205/809/1/L5}, 809

\bibitem[\protect\citeauthoryear{Dong, Li, Chiang  \& Li}{Dong
  et~al.}{2017}]{Dong17}
Dong R.,  Li S.,  Chiang E.,   Li H.,  2017, \mn@doi [The Astrophysical
  Journal] {10.3847/1538-4357/aa72f2}, 843, 127

\bibitem[\protect\citeauthoryear{Draine}{Draine}{2006}]{Draine2006}
Draine B.~T.,  2006, \mn@doi [The Astrophysical Journal] {10.1086/498130}, 636,
  1114

\bibitem[\protect\citeauthoryear{Dr{\c{a}}{\.{z}}kowska, Li, Birnstiel,
  Stammler  \& Li}{Dr{\c{a}}{\.{z}}kowska et~al.}{2019}]{Drazkowska2019}
Dr{\c{a}}{\.{z}}kowska J.,  Li S.,  Birnstiel T.,  Stammler S.~M.,   Li H.,
  2019, \mn@doi [The Astrophysical Journal] {10.3847/1538-4357/ab46b7}, 885, 91

\bibitem[\protect\citeauthoryear{Dubrulle, Morfill  \& Sterzik}{Dubrulle
  et~al.}{1995}]{Dubrulle1995}
Dubrulle B.,  Morfill G.,   Sterzik M.,  1995, \mn@doi [Icarus]
  {10.1006/icar.1995.1058}, 114, 237

\bibitem[\protect\citeauthoryear{Dullemond, Juhasz, Sereshti, Peters, Commercon
   \& Flock}{Dullemond et~al.}{2012}]{Dullemond2012}
Dullemond C.~P.,  Juhasz A.;~Pohl A.,  Sereshti F.;~Shetty R.,  Peters T.,
  Commercon B.,   Flock M.,  2012, {RADMC-3D: A multi-purpose radiative
  transfer tool}

\bibitem[\protect\citeauthoryear{D{\"{u}}rmann \& Kley}{D{\"{u}}rmann \&
  Kley}{2015}]{Durmann15}
D{\"{u}}rmann C.,  Kley W.,  2015, \mn@doi [Astronomy and Astrophysics]
  {10.1051/0004-6361/201424837}, 574

\bibitem[\protect\citeauthoryear{Edgar \& Quillen}{Edgar \&
  Quillen}{2008}]{Edgar2008}
Edgar R.~G.,  Quillen A.~C.,  2008, \mn@doi [Monthly Notices of the Royal
  Astronomical Society] {10.1111/j.1365-2966.2008.13242.x}, 387, 387

\bibitem[\protect\citeauthoryear{Fedele et~al.,}{Fedele
  et~al.}{2017}]{Fedele2017}
Fedele D.,  et~al., 2017, \mn@doi [Astronomy {\&} Astrophysics]
  {10.1051/0004-6361/201629860}, 600, A72

\bibitem[\protect\citeauthoryear{Fouchet, Maddison, Gonzalez, Murray, Fouchet,
  Maddison, Gonzalez  \& Murray}{Fouchet et~al.}{2007}]{Fouchet2007}
Fouchet L.,  Maddison S.,  Gonzalez J.-F.,  Murray J.,  Fouchet L.,  Maddison
  S.~T.,  Gonzalez J.-f.,   Murray J.~R.,  2007, Astronomy {\&} Astrophysics,
  474, 1037–1047

\bibitem[\protect\citeauthoryear{Fouchet, Gonzalez  \& Maddison}{Fouchet
  et~al.}{2010}]{Fouchet2010}
Fouchet L.,  Gonzalez J.-F.,   Maddison S.~T.,  2010, \mn@doi [Astronomy {\&}
  Astrophysics] {10.1051/0004-6361/201218806}, 518, A16

\bibitem[\protect\citeauthoryear{Fromang \& Nelson}{Fromang \&
  Nelson}{2009}]{Fromang2009}
Fromang S.,  Nelson R.~P.,  2009, \mn@doi [Astronomy and Astrophysics]
  {10.1051/0004-6361/200811220}, 496, 597

\bibitem[\protect\citeauthoryear{Fung \& Chiang}{Fung \&
  Chiang}{2016}]{Fung2016}
Fung J.,  Chiang E.,  2016, \mn@doi [The Astrophysical Journal]
  {10.3847/0004-637x/832/2/105}, 832, 105

\bibitem[\protect\citeauthoryear{Goldreich \& Tremaine}{Goldreich \&
  Tremaine}{1980}]{GoldreichTremaine1980}
Goldreich P.,  Tremaine S.,  1980, The Astrophysical Journal, 241, 425

\bibitem[\protect\citeauthoryear{Gonzalez, Pinte, Maddison  \&
  M{\'{e}}nard}{Gonzalez et~al.}{2012}]{Gonzalez2012}
Gonzalez J.~F.,  Pinte C.,  Maddison S.~T.,   M{\'{e}}nard F.,  2012, \mn@doi
  [Astronomy {\&} Astrophysics] {10.1017/S1743921313008053}, 547, A58

\bibitem[\protect\citeauthoryear{Gonzalez, Laibe, Maddison, Pinte  \&
  M{\'{e}}nard}{Gonzalez et~al.}{2015}]{Gonzalez2015}
Gonzalez J.~F.,  Laibe G.,  Maddison S.~T.,  Pinte C.,   M{\'{e}}nard F.,
  2015, \mn@doi [Monthly Notices of the Royal Astronomical Society: Letters]
  {10.1093/mnrasl/slv120}, 454, L36

\bibitem[\protect\citeauthoryear{Hildebrand}{Hildebrand}{1983}]{Hildebrand1983}
Hildebrand R.~H.,  1983, Quarterly Journal of the Royal Astronomical Society,
  pp 267--282

\bibitem[\protect\citeauthoryear{Humphries \& Nayakshin}{Humphries \&
  Nayakshin}{2018}]{Humphries2018}
Humphries J.,  Nayakshin S.,  2018, \mn@doi [Monthly Notices of the Royal
  Astronomical Society] {10.1093/mnras/stz2497}, 477, 593

\bibitem[\protect\citeauthoryear{Humphries \& Nayakshin}{Humphries \&
  Nayakshin}{2019}]{Humphries2019}
Humphries J.,  Nayakshin S.,  2019, Monthly Notices of the Royal Astronomical
  Society, 489, 5187

\bibitem[\protect\citeauthoryear{Isella \& Turner}{Isella \&
  Turner}{2018}]{IsellaTurner2018}
Isella A.,  Turner N.~J.,  2018, \mn@doi [The Astrophysical Journal]
  {10.3847/1538-4357/aabb07}, 860, 27

\bibitem[\protect\citeauthoryear{Jin, Li, Isella, Li  \& Ji}{Jin
  et~al.}{2016}]{Jin2016}
Jin S.,  Li S.,  Isella A.,  Li H.,   Ji J.,  2016, \mn@doi [The Astrophysical
  Journal] {10.3847/0004-637x/818/1/76}, 818, 76

\bibitem[\protect\citeauthoryear{Johansen, Youdin  \& Klahr}{Johansen
  et~al.}{2009}]{Johansen2009}
Johansen A.,  Youdin A.,   Klahr H.,  2009, \mn@doi [Astrophysical Journal]
  {10.1088/0004-637X/697/2/1269}, 697, 1269

\bibitem[\protect\citeauthoryear{Kanagawa, Muto, Tanaka, Tanigawa, Takeuchi,
  Tsukagoshi  \& Momose}{Kanagawa et~al.}{2016}]{Kanagawa2016MassWidth}
Kanagawa K.~D.,  Muto T.,  Tanaka H.,  Tanigawa T.,  Takeuchi T.,  Tsukagoshi
  T.,   Momose M.,  2016, \mn@doi [Publications of the Astronomical Society of
  Japan] {10.1093/pasj/psw037}, 68, 1

\bibitem[\protect\citeauthoryear{Kanagawa, Tanaka, Muto  \& Tanigawa}{Kanagawa
  et~al.}{2017}]{Kanagawa2017}
Kanagawa K.~D.,  Tanaka H.,  Muto T.,   Tanigawa T.,  2017, \mn@doi
  [Publications of the Astronomical Society of Japan] {10.1093/pasj/psx114},
  69, 97

\bibitem[\protect\citeauthoryear{Keppler et~al.,}{Keppler
  et~al.}{2018}]{Keppler2018}
Keppler M.,  et~al., 2018, \mn@doi [Astronomy and Astrophysics]
  {10.1051/0004-6361/201832957}, 617, A44
  
\bibitem[\protect\citeauthoryear{Kley \& Dirksen}{Kley \& Dirksen}{2006}]{Kley2006}
Kley, W. \& Dirksen, G.,  2006, \mn@doi [Astronomy and Astrophysics]
  {10.1051/0004-6361:20053914}, 447, 1, 369-377

\bibitem[\protect\citeauthoryear{Kley \& Nelson}{Kley \&
  Nelson}{2012}]{Kley2012}
Kley W.,  Nelson R.~P.,  2012, \mn@doi [Annual Review of Astronomy and
  Astrophysics] {10.1146/annurev-astro-081811-125523}, 50, 211

\bibitem[\protect\citeauthoryear{Laibe \& Price}{Laibe \&
  Price}{2012}]{Laibe2012}
Laibe G.,  Price D.~J.,  2012, \mn@doi [Monthly Notices of the Royal
  Astronomical Society] {10.1111/j.1365-2966.2011.20201.x}, 420, 2365

\bibitem[\protect\citeauthoryear{LeVeque}{LeVeque}{2002}]{LeVeque2010}
LeVeque R.~J.,  2002, {Finite Volume Methods for Hyperbolic Problems}.
Cambridge University Press

\bibitem[\protect\citeauthoryear{LeVeque}{LeVeque}{2004}]{LeVeque2004}
LeVeque R.~J.,  2004, \mn@doi [Journal of Hyperbolic Differential Equations]
  {10.1142/s0219891604000135}, 01, 315

\bibitem[\protect\citeauthoryear{Li \& Greenberg}{Li \&
  Greenberg}{1997}]{Li1997}
Li A.,  Greenberg J.~M.,  1997, Astronomy {\&} Astrophysics, 323, 566

\bibitem[\protect\citeauthoryear{Lin \& Papaloizou}{Lin \&
  Papaloizou}{1984}]{Lin1984}
Lin D. N.~C.,  Papaloizou J. C.~B.,  1984, The Astrophysical Journal, 285, 818

\bibitem[\protect\citeauthoryear{Lin \& Papaloizou}{Lin \&
  Papaloizou}{1986}]{LinPapa1986}
Lin D. N.~C.,  Papaloizou J.,  1986, \mn@doi [The Astrophysical Journal]
  {10.1086/164653}, 309, 846

\bibitem[\protect\citeauthoryear{Lin \& Papaloizou}{Lin \&
  Papaloizou}{1993}]{Lin1993}
Lin D. N.~C.,  Papaloizou J. C.~B.,  1993, in Levy E.,  Lunine J.,  eds, ,
  Protostars and Planets III.
University of Arizona Press, p.~749

\bibitem[\protect\citeauthoryear{Liu}{Liu}{2019}]{Liu2019}
Liu H.~B.,  2019, \mn@doi [The Astrophysical Journal]
  {10.3847/2041-8213/ab1f8e}, 877, L22

\bibitem[\protect\citeauthoryear{Lynden-Bell \& Pringle}{Lynden-Bell \&
  Pringle}{1974}]{Lynden-Bell1974}
Lynden-Bell D.,  Pringle J.~E.,  1974, \mn@doi [Monthly Notices of the Royal
  Astronomical Society] {10.1093/mnras/168.3.603}, 168, 603
  
  
\bibitem[\protect\citeauthoryear{Masset \& Papaloizou}{Masset \& Papaloizou}{2003}]{Masset2003}
Masset, F.~S., Papaloizou, J.~C.~B.,  2003, \mn@doi [The Astrophysical
  Journal] {10.1086/373892}, 588, 1, 494-508

\bibitem[\protect\citeauthoryear{Mcmullin, Waters, Schiebel, Young  \&
  Golap}{Mcmullin et~al.}{2007}]{CASA2007}
Mcmullin J.~P.,  Waters B.,  Schiebel D.,  Young W.,   Golap K.,  2007,
  Astronomical Data Analysis Software and Systems XVI, 376, 127

\bibitem[\protect\citeauthoryear{M{\"{u}}ller et~al.,}{M{\"{u}}ller
  et~al.}{2018}]{Mueller2018}
M{\"{u}}ller A.,  et~al., 2018, \mn@doi [Astronomy {\&} Astrophysics]
  {10.1051/0004-6361/201833584}, 617, L2

\bibitem[\protect\citeauthoryear{Nakagawa, Sekiya  \& Hayashi}{Nakagawa
  et~al.}{1986}]{Nakagawa1986}
Nakagawa Y.,  Sekiya M.,   Hayashi C.,  1986, ICARUS, 67, 375

\bibitem[\protect\citeauthoryear{Paardekooper \& Mellema}{Paardekooper \&
  Mellema}{2004}]{Paardekooper2004}
Paardekooper S.-J.,  Mellema G.,  2004, Astronomy {\&} Astrophysics, 425,
  L9–L12

\bibitem[\protect\citeauthoryear{Paardekooper \& Mellema}{Paardekooper \&
  Mellema}{2006}]{Paardekooper2006}
Paardekooper S.-J.,  Mellema G.,  2006, \mn@doi [Astronomy {\&} Astrophysics]
  {10.1051/0004-6361:20054449}, 453, 1129

\bibitem[\protect\citeauthoryear{Perez, Dunhill, Casassus, Roman,
  Szul{\'{a}}gyi, Flores, Marino  \& Montesinos}{Perez
  et~al.}{2015}]{Perez2015}
Perez S.,  Dunhill A.,  Casassus S.,  Roman P.,  Szul{\'{a}}gyi J.,  Flores C.,
   Marino S.,   Montesinos M.,  2015, \mn@doi [Astrophysical Journal Letters]
  {10.1088/2041-8205/811/1/L5}, 811, L5

\bibitem[\protect\citeauthoryear{Picogna \& Kley}{Picogna \&
  Kley}{2015}]{Picogna2015}
Picogna G.,  Kley W.,  2015, \mn@doi [Astronomy {\&} Astrophysics]
  {10.1051/0004-6361/201526921}, 584, A110

\bibitem[\protect\citeauthoryear{Pineda et~al.,}{Pineda
  et~al.}{2019}]{Pineda2019}
Pineda J.~E.,  et~al., 2019, \mn@doi [ApJ] {10.3847/1538-4357/aaf389}, 871, 48

\bibitem[\protect\citeauthoryear{Pinilla, Benisty  \& Birnstiel}{Pinilla
  et~al.}{2012}]{Pinilla2012}
Pinilla P.,  Benisty M.,   Birnstiel T.,  2012, \mn@doi [Astronomy and
  Astrophysics] {10.1051/0004-6361/201219315}, 545, 81
  
 \bibitem[\protect\citeauthoryear{Pinte et~al.,}{Pinte et~al.}{2016}]{Pinte2016}
 Pinte C.,  et~al.,  2016, \mn@doi [The Astrophysical Journal]
  {10.3847/0004-637X}, 816, 1, 25

\bibitem[\protect\citeauthoryear{Pinte et~al.,}{Pinte et~al.}{2018}]{Pinte2018}
Pinte C.,  et~al., 2018, \mn@doi [The Astrophysical Journal Letters]
  {10.3847/2041-8213/aac6dc}, 860, L13

\bibitem[\protect\citeauthoryear{Rice, Armitage, Wood  \& Lodato}{Rice
  et~al.}{2006}]{Rice2006}
Rice W.~K.,  Armitage P.~J.,  Wood K.,   Lodato G.,  2006, \mn@doi [Monthly
  Notices of the Royal Astronomical Society]
  {10.1111/j.1365-2966.2006.11113.x}, 373, 1619

\bibitem[\protect\citeauthoryear{Robert, Crida, Lega, M{\'{e}}heut  \&
  Morbidelli}{Robert et~al.}{2018}]{Robert18}
Robert C.~M.,  Crida A.,  Lega E.,  M{\'{e}}heut H.,   Morbidelli A.,  2018,
  \mn@doi [Astronomy and Astrophysics] {10.1051/0004-6361/201833539}, 617

\bibitem[\protect\citeauthoryear{Rosotti, Juhasz, Booth  \& Clarke}{Rosotti
  et~al.}{2016}]{Rosotti2016}
Rosotti G.~P.,  Juhasz A.,  Booth R.~A.,   Clarke C.~J.,  2016, \mn@doi
  [Monthly Notices of the Royal Astronomical Society] {10.1093/mnras/stw691},
  459, 2790

\bibitem[\protect\citeauthoryear{Ruge, Flock, Wolf, Dzyurkevich, Fromang,
  Henning, Klahr  \& Meheut}{Ruge et~al.}{2016}]{Ruge2016}
Ruge J.~P.,  Flock M.,  Wolf S.,  Dzyurkevich N.,  Fromang S.,  Henning T.,
  Klahr H.,   Meheut H.,  2016, \mn@doi [Astronomy and Astrophysics]
  {10.1051/0004-6361/201526616}, 590

\bibitem[\protect\citeauthoryear{Rybicki \& Lightman}{Rybicki \&
  Lightman}{1979}]{Rybicki1979}
Rybicki G.~B.,  Lightman A.~P.,  1979, {Radiative processes in astrophysics}.
Wiley

\bibitem[\protect\citeauthoryear{Shakura \& Sunyaev}{Shakura \&
  Sunyaev}{1973}]{Shakura1973}
Shakura N.~I.,  Sunyaev R.~A.,  1973, \mn@doi [Symposium - International
  Astronomical Union] {10.1017/s007418090010035x}, 55, 155

\bibitem[\protect\citeauthoryear{Stone}{Stone}{1997}]{Stone1997}
Stone J.~M.,  1997, \mn@doi [The Astrophysical Journal] {10.1086/304595}, 487,
  271

\bibitem[\protect\citeauthoryear{Szul{\'{a}}gyi}{Szul{\'{a}}gyi}{2017}]{Szulagyi2017}
Szul{\'{a}}gyi J.,  2017, \mn@doi [The Astrophysical Journal]
  {10.3847/1538-4357/aa7515}, 842, 103

\bibitem[\protect\citeauthoryear{Szul{\'{a}}gyi \& Garufi}{Szul{\'{a}}gyi \&
  Garufi}{2019}]{Szulagyi2019}
Szul{\'{a}}gyi J.,  Garufi A.,  2019, arXiv e-prints, arXiv:1906

\bibitem[\protect\citeauthoryear{Szul{\'{a}}gyi, Morbidelli, Crida  \&
  Masset}{Szul{\'{a}}gyi et~al.}{2014}]{Szulagyi2014}
Szul{\'{a}}gyi J.,  Morbidelli A.,  Crida A.,   Masset F.,  2014, \mn@doi
  [Astrophysical Journal] {10.1088/0004-637X/782/2/65}, 782

\bibitem[\protect\citeauthoryear{Szul{\'{a}}gyi, Masset, Lega, Crida,
  Morbidelli  \& Guillot}{Szul{\'{a}}gyi et~al.}{2016}]{Szulagyi2016}
Szul{\'{a}}gyi J.,  Masset F.,  Lega E.,  Crida A.,  Morbidelli A.,   Guillot
  T.,  2016, \mn@doi [Monthly Notices of the Royal Astronomical Society]
  {10.1093/mnras/stw1160}, 460, 2853

\bibitem[\protect\citeauthoryear{Szul{\'{a}}gyi, van~der Plas, Meyer, Pohl,
  Quanz, Mayer, Daemgen  \& Tamburello}{Szul{\'{a}}gyi
  et~al.}{2018a}]{Szulagyi2018b}
Szul{\'{a}}gyi J.,  van~der Plas G.,  Meyer M.~R.,  Pohl A.,  Quanz S.~P.,
  Mayer L.,  Daemgen S.,   Tamburello V.,  2018a, \mn@doi [Monthly Notices of
  the Royal Astronomical Society] {10.1093/mnras/stx2602}, 473, 3573

\bibitem[\protect\citeauthoryear{Szul{\'{a}}gyi, van~der Plas, Meyer, Pohl,
  Quanz, Mayer, Daemgen  \& Tamburello}{Szul{\'{a}}gyi
  et~al.}{2018b}]{Szulagyi2018}
Szul{\'{a}}gyi J.,  van~der Plas G.,  Meyer M.~R.,  Pohl A.,  Quanz S.~P.,
  Mayer L.,  Daemgen S.,   Tamburello V.,  2018b, \mn@doi [Monthly Notices of
  the Royal Astronomical Society] {10.1093/mnras/stx2602}, 473, 3573

\bibitem[\protect\citeauthoryear{Szul{\'{a}}gyi, Dullemond, Pohl  \&
  Quanz}{Szul{\'{a}}gyi et~al.}{2019}]{Szulagyi2018a}
Szul{\'{a}}gyi J.,  Dullemond C.~P.,  Pohl A.,   Quanz S.~P.,  2019, Monthly
  Notices of the Royal Astronomical Society, 487, 1248

\bibitem[\protect\citeauthoryear{Szul{\'{a}}gyi, Binkert  \&
  Surville}{Szul{\'{a}}gyi et~al.}{2021}]{Szulagyi2021}
Szul{\'{a}}gyi J.,  Binkert F.,   Surville C.,  2021, arXiv e-prints

\bibitem[\protect\citeauthoryear{Tanaka, Takeuchi  \& Ward}{Tanaka
  et~al.}{2002}]{Tanaka2002}
Tanaka H.,  Takeuchi T.,   Ward W.~R.,  2002, \mn@doi [The Astrophysical
  Journal] {10.1086/380992}, 565, 1257

\bibitem[\protect\citeauthoryear{Teague, Bae, Bergin, Birnstiel  \&
  Foreman-Mackey}{Teague et~al.}{2018}]{Teague2018}
Teague R.,  Bae J.,  Bergin E.~A.,  Birnstiel T.,   Foreman-Mackey D.,  2018,
  \mn@doi [The Astrophysical Journal Letters] {10.3847/2041-8213/aac6d7}, 860,
  L12

\bibitem[\protect\citeauthoryear{Toomre}{Toomre}{1964}]{Toomre1964}
Toomre A.,  1964, \mn@doi [The Astrophysical Journal]
  {10.3847/0004-637x/832/2/201}, 129, 1217

\bibitem[\protect\citeauthoryear{Toro}{Toro}{2009}]{Toro2009}
Toro E.~F.,  2009, {Riemann solvers and numerical methods for fluid dynamics: A
  practical introduction}.
Springer

\bibitem[\protect\citeauthoryear{Tychoniec et~al.,}{Tychoniec
  et~al.}{2020}]{Tychoneiec2020}
Tychoniec L.,  et~al., 2020, \mn@doi [Astronomy {\&} Astrophysics]
  {10.1051/0004-6361/202037851}, 640, A19

\bibitem[\protect\citeauthoryear{Van Der~Marel et~al.,}{Van Der~Marel
  et~al.}{2013}]{vandermarel2013}
Van Der~Marel N.,  et~al., 2013, \mn@doi [Science] {10.1126/science.1236770},
  340, 1199

\bibitem[\protect\citeauthoryear{Vorobyov, Matsukoba, Omukai  \&
  Guedel}{Vorobyov et~al.}{2020}]{Voroboyov2020}
Vorobyov E.~I.,  Matsukoba R.,  Omukai K.,   Guedel M.,  2020, \mn@doi
  [Astronomy {\&} Astrophysics] {10.1051/0004-6361/202037841}, 638, A102

\bibitem[\protect\citeauthoryear{Weber, Ben{\'{i}}tez-Llambay, Gressel, Krapp
  \& Pessah}{Weber et~al.}{2018}]{Weber2018}
Weber P.,  Ben{\'{i}}tez-Llambay P.,  Gressel O.,  Krapp L.,   Pessah M.~E.,
  2018, \mn@doi [The Astrophysical Journal] {10.3847/1538-4357/aaab63}, 854,
  153

\bibitem[\protect\citeauthoryear{Weber, Sebasti´ Sebasti´an P ´~Erez,
  Ben´itezben´itez-Llambay, Gressel, Casassus  \& Krapp}{Weber
  et~al.}{2019}]{Weber20}
Weber P.,  Sebasti´ Sebasti´an P ´~Erez S.,  Ben´itezben´itez-Llambay P.,
  Gressel O.,  Casassus S.,   Krapp L.,  2019, The Astrophysical Journal, 884,
  178

\bibitem[\protect\citeauthoryear{Weidenschilling}{Weidenschilling}{1977}]{Weidenschilling1977}
Weidenschilling S.~J.,  1977, Monthly Notices of the Royal Astronomical
  Society, 180, 57

\bibitem[\protect\citeauthoryear{Whipple}{Whipple}{1972}]{Whipple1972}
Whipple F.~L.,  1972, From Plasma to Planet, Proceedings of the Twenty-First
  Nobel Symposium held 6-10 September, 1971 at Saltsj{\"{o}}baden, near
  Stockholm, Sweden., p.~211

\bibitem[\protect\citeauthoryear{Youdin \& Lithwick}{Youdin \&
  Lithwick}{2007}]{Youdin2007}
Youdin A.~N.,  Lithwick Y.,  2007, \mn@doi [Icarus]
  {10.1016/j.icarus.2007.07.012}, 192, 588

\bibitem[\protect\citeauthoryear{Zhang \& Zhu}{Zhang \&
  Zhu}{2020}]{ZhangZhu2020}
Zhang S.,  Zhu Z.,  2020, \mn@doi [Monthly Notices of the Royal Astronomical
  Society] {10.1093/mnras/staa404}, 493, 2287

\bibitem[\protect\citeauthoryear{Zhang, Blake  \& Bergin}{Zhang
  et~al.}{2015}]{Zhang2015}
Zhang K.,  Blake G.~A.,   Bergin E.~A.,  2015, \mn@doi [Astrophysical Journal
  Letters] {10.1088/2041-8205/806/1/L7}, 806

\bibitem[\protect\citeauthoryear{Zhang et~al.,}{Zhang et~al.}{2018}]{Zhang2018}
Zhang S.,  et~al., 2018, \mn@doi [The Astrophysical Journal Letters]
  {10.3847/2041-8213/aaf744}, 869, L47

\bibitem[\protect\citeauthoryear{Zhu, Nelson, Dong, Espaillat  \& Hartmann}{Zhu
  et~al.}{2012}]{Zhu2012}
Zhu Z.,  Nelson R.~P.,  Dong R.,  Espaillat C.,   Hartmann L.,  2012, \mn@doi
  [Astrophysical Journal] {10.1088/0004-637X/755/1/6}, 755

\bibitem[\protect\citeauthoryear{Zhu et~al.,}{Zhu et~al.}{2019}]{Zhu2019b}
Zhu Z.,  et~al., 2019, \mn@doi [The Astrophysical Journal]
  {10.3847/2041-8213/ab1f8c}, 877, L18

\bibitem[\protect\citeauthoryear{Ziampras, Kley  \& Dullemond}{Ziampras
  et~al.}{2020}]{Ziampras2020}
Ziampras A.,  Kley W.,   Dullemond C.~P.,  2020, Astronomy {\&} Astrophysics,
  637, A50

\bibitem[\protect\citeauthoryear{Zubko, Krelowski  \& Wegner}{Zubko
  et~al.}{1996}]{Zubko1996}
Zubko V.~G.,  Krelowski J.,   Wegner W.,  1996, Monthly Notices of the Royal
  Astronomical Society, 283, 577

\bibitem[\protect\citeauthoryear{de Val-Borro et~al.,}{de~Val-Borro
  et~al.}{2006}]{DeVal-Borro2006}
de Val-Borro M.,  et~al., 2006, Monthly Notices of the Royal Astronomical
  Society, 370, 529–558

\makeatother
\end{thebibliography}





\bsp	
\label{lastpage}
\end{document}